\documentclass[letterpaper,aps,prd,twocolumn,tightenlines,preprintnumbers,nofootinbib,showkeys,superscriptaddress,longbibliography
]{revtex4-1}

\pdfoutput=1 

\usepackage[T1]{fontenc}

\usepackage{fullpage}
\usepackage{amsfonts}
\usepackage{amsmath}
\usepackage{slashed}
\usepackage{amssymb}
\usepackage{latexsym}
\usepackage[dvipsnames]{xcolor}
\usepackage{float}
\usepackage{multirow}
\usepackage{lipsum} 
\usepackage{url}
\usepackage[breaklinks=true]{hyperref}
\usepackage{enumitem}
\hypersetup{colorlinks=true,citecolor=blue,linkcolor=blue,urlcolor=NavyBlue}
\usepackage[caption=false]{subfig}
\usepackage{natbib}
\usepackage{cancel}
\usepackage{relsize}
\usepackage[left=2cm,right=2cm,top=2cm,bottom=2cm]{geometry}
\usepackage{diagbox}
\usepackage{makecell}
\usepackage{soul}
\usepackage{mathtools} 
\usepackage{feynmp-auto} 
\usepackage{ulem} 

\usepackage{bm}
\usepackage{subfig}

\usepackage{shorthand}
\usepackage{mciteplus}

\def\beq{\begin{equation}}
\def\eeq{\end{equation}}
\def\bea{\begin{eqnarray}}
\def\eea{\end{eqnarray}}

\def\h1{\ensuremath{h_1}}
\def\h2{\ensuremath{h_2}}

\begin{document}
\linespread{1.02}

\title{Confront a dilaton model with the LHC measurements}


	\author{J.E. Wu}
	\email{wujinger22@mails.ucas.ac.cn}
	\affiliation{School of Physics Sciences, University of Chinese Academy of Sciences, Beijing 100039, P.R. China.\\}

    \author{Q.S. Yan}
    \email{yanqishu@ucas.ac.cn}
	\affiliation{Center for Future High Energy Physics, Chinese Academy of Sciences, Beijing 100049, P.R. China.}
	\affiliation{School of Physics Sciences, University of Chinese Academy of Sciences, Beijing 100039, P.R. China.\\}

\begin{abstract}
The origin of the Higgs boson ($H_{125}$), discovered in 2012, remains a mystery. In the metric affine theory (MAT) framework, we study the scalar potential and investigate a couple of scenarios for the symmetry breaking mechanisms with a dilaton model which is derived from the geometry. The LHC constraints for the couplings of Yukawa couplings, Higgs-weak vector bosons and Higgs self-couplings, in this model are examined, which identify the parameter space where the discovered Higgs boson $m_h=125$ GeV can be dilaton-dominant and the features of Higgs self-couplings are explored. It is found that via the measurements of Higgs pair production, the High Luminosity LHC (HL-LHC) running can either confirm or rule out the dilaton dominance.
\end{abstract}


\maketitle
\section{Introduction}
Great progress has been achieved recently for the Standard Model (SM) and General Relativity (GR), which are both successful theories for interpreting fundamental interactions in nature. The SM has worked quite well in accommodating high-energy data. The Higgs discovery in 2012 is a great leap for the development of particle physics \cite{ATLAS:2012yve,CMS:2012qbp}, which has found the last piece of the SM, $H_{125}$, and has further confirmed the pillars of the SM, such as the concepts of non-Abelian local gauge theory (Yang-Mills theory), the spontaneous symmetry breaking (SSB), and fermion mass generation via Yukawa type couplings. The GR is applicable to physics on large scales and its prediction is confirmed directly by the first observation of the gravitational wave in 2016 \cite{LIGOScientific:2016aoc}. However, the renormalizability of GR obstructs the quantization of gravity, which is required for the unification of all interactions. Then, it motivates various models to modify gravity \cite{Stelle:1976gc,Horava:2009uw,Modesto:2011kw}.
Among these models, we are interested in the branch of thoughts which uses scale symmetry to connect the GR with the SM.
The localization of the scale symmetry \cite{hayashi_elementary_1977,hayashi_remarks_1979,ghilencea_Weyl_2019,ghilencea_spontaneous_2019,ghilencea_stueckelberg_2020,ghilencea_standard_2021,ghilencea_non-metric_2023,ghilencea_standard_2024}, which is also called Weyl symmetry or conformal symmetry introduced below, is possible to
construct a ultraviolet (UV) complete model.

The concept of gauged scale symmetry was originated from Weyl's work, which was aimed at unifying gravity and electromagnetic theory by introducing a Weyl vector boson to play the role of Maxwell vector field \cite{Weyl:1918ib}. Due to the fact that it contradicted with the atomic data, the idea was abandoned for a long time. It was revived by Dirac \cite{Dirac:1973gk} in order to understand the physical laws between cosmological scale and atomic scale. Thus it enlighten a wealth of research for gauge unification theory with the Weyl symmetry
\cite{ORaifeartaigh:1997dvq,Fujii:1982ms,Drechsler:1998gy,Chamseddine:2006ep}. 

The global scale symmetry and its breaking are important guiding principles to formulate its effective Lagrangian and the interaction between the Weyl vector and the Standard Model. 
The Higgs potential of the SM can be given as 
\begin{equation}
V(\phi) = -\mu^2 \phi^\dagger \phi+ \lambda (\phi^\dagger \phi)^2\,,
\label{V_SM}
\end{equation}
where $\lambda$ is the self-couplings of Higgs boson and $\mu$ is a mass parameter which breaks the scale symmetry. The SM preserves the 
scale symmetry if the parameter $\mu \to 0$. As it is well-known that the scale symmetry of fundamental dynamics might related to the dilaton (a singlet field) \cite{Kaluza:1921tu,Klein:1926tv,Becker:2006dvp}, which can be the corresponding Nambu Goldstone (NG) boson as a massless particle near the electroweak scale. It is natural to conjecture that $H_{125}$ might be related to the dilaton, such as the CFT dilaton \cite{Barger_2012}, the techni-dilaton (TD) \cite{matsuzaki_discovering_2012}, or, more generally, the global scale transformation dilaton \cite{Bellazzini_2013, Abe:2012eu}. However, the gauged scale transformation dilaton has been less discussed.

An early attempt which associates Higgs with dilaton giving a gravitational origin for Higgs is shown in \cite{Flato:1987bb}. In the reference \cite{vanderBij:1993hx}, the importance of the large non-minimal coupling constant $\xi$ was emphasized and Higgs field was regarded as a physical degree of freedom. Motivated by the great success of gauge theories, Weyl vector was introduced into electroweak (EW) interaction by Cheng \cite{PhysRevLett.61.2182} through the application of gauged scale symmetry, where the Higgs field was assumed to play the role of Goldstone particle which can be eaten by Weyl vector. In later literature \cite{Nishino:2004kb,Nishino:2009in}, one more scalar field was introduced and Higgs boson can be a physical degree of freedom.  The gravitational origin can provide a new perspective on interpreting the SM through an inherent geometry framework \cite{deCesare:2016mml}, which is called Weyl geometry from the prospect of the metric affine theory (MAT) framework \cite{hehl_metric_1995,Dirac:1973gk,trautman_geometry_1979}.  

The referenced studies rely on linear gravity theory, whereas the more natural approach is quadratic gravity within Weyl’s geometric framework. 
Recently a few works \cite{ghilencea_standard_2021,ghilencea_standard_2024} have applied it to the SM, which is called as the SMW. The SMW is an interesting model which can not only unify the gravity and the SM, but also can have successful inflation, similar to the Starobinsky $R^2$-inflation.
It is curious to us whether the combination of Weyl geometry and SM can accommodate the LHC data, especially the Higgs data, collected over more than a decade \cite{ATLAS:2022vkf,CMS:2012qbp}. To our best knowledge, such a question has not been explored in literature yet. This work is supposed to fill this gap.

In this work, we consider a dilaton model, in which quadratic gravity is embedded in the SM within the Weyl geometry.
In the weak eigenstates, there are two scalar fields in the model. The first one is a singlet $\Phi$, which is called as dilaton, and the second one is a doublet $\phi$.
The $H_{125}$ is parametrized in terms of dilaton and the doublet fields more comprehensively, which yields two different scenarios and can have significantly distinct properties as shown in this work. It is found that our model can accommodate the LHC data, which might be a key to address some BSM issues, such as whether the $H_{125}$ is fundamental or composite, as pointed out by \cite{Degrassi:2016wml,Lane:2022ybv,Steingasser:2023ugv}.

Specifically, the Higgs potentials of the model are confronted with the $\kappa$-framework using the latest experimental measurements \cite{ATLAS:2024ish} after spontaneous symmetry breaking of the local scale and electroweak symmetry. We identify the region of parameter space where $H_{125}$ is dilaton-dominant and investigate the capability of the High-Luminosity Large Hadron Collider (HL-LHC) \cite{Dainese:2019rgk} to probe this region. It is found that HL-LHC can confirm/rule out the dilaton dominance.

This paper is organized as given below. In Section II, a brief introduction to the dilaton model is given. In Section III, two major scenarios of the model are presented. In Section IV, a numerical analysis on the parameter space of these two scenarios is provided. In Section V, scalar potentials in literature are compared and the electroweak spontaneous symmetry breaking mechanism is examined. Finally, several discussion and conclusions are provided.

\section{A brief introduction to the dilaton model}
The similarity of Yang-Mills gauge fields and the affine connections provides a framework to unify the SM and the gravity through the construction of covariant derivative within the perspective of geometry, which can be generalized to the Weyl conformal geometry (indicated by a hat symbol) after extending metricity to the non-metricity property $\hat \nabla_\mu g_{\rho \sigma} = -2 \omega_\mu g_{\rho \sigma}$ \cite{hehl_metric_1995}. 
The dilaton model includes a singlet real scalar field $\Phi$ and a complex doublet scalar for field $\phi$, then the non-minimal couplings between scalar fields and Ricci scalar are also included to unify the SM and gravity. The whole Lagrangian can be organized as bosonic and fermionic parts as given below
\begin{equation}
     \begin{aligned}[t]
\sqrt{-g} {\cal L} =\,& \sqrt{-g} \left( {\cal L}_B + {\cal L}_F \right )\,.
\end{aligned}
 \label{L_Orig}
\end{equation}
where the bosonic part can be written as given below
\begin{equation}
     \begin{aligned}[t]
{\cal L}_B = & {\cal L}_{Bk} +  {\cal L}_V \,,\\
{\cal L}_{Bk} =\,& \frac{\hat R^2}{4!\xi^2} -\frac{\hat C_{\mu\nu\rho\sigma}^2}{\eta^2} - \frac{1}{4 g_w^2} \hat W_{\mu\nu}^a \hat W^{\mu\nu,a} - \frac{1}{4 g_s^2} \hat F_{\mu\nu}\hat F^{\mu\nu}\\
& +\frac{1}{2}  \hat \nabla_\mu \Phi \hat \nabla^\mu \Phi + \hat \nabla_\mu \phi^\dagger  \hat \nabla^\mu \phi \,,  \\
 - {\cal L}_V  =\,& \frac{\rho}{4!}\Phi^4 + \frac{\alpha}{2}\Phi^2 \phi^\dagger \phi  + \frac{\lambda}{4} (\phi^\dagger \phi)^2 + \frac{\beta}{2}\Phi^2 \hat R + \gamma \phi^\dagger \phi \hat R \,.
 \end{aligned}
\label{L_Orig1}
\end{equation}
In the kinetic terms of the bosonic part, four couplings, $\xi$, $\eta$, $g_w$, and $g_s$, are introduced because of the inclusion of the Ricci scalar, the Riemann tensor, the SM gauge bosons, and the Weyl vector. Note that although a relation between $\hat{C}_{\mu \nu \rho \sigma}^2$ and $\hat{F}_{\mu \nu} \hat{F}^{\mu \nu}$ is shown in Eq.~\eqref{W2R}, both of them are independent and conformally invariant, as discussed in Appendix A of Ref.~\cite{Drechsler:1998gy}. In addition to the three free parameters $\rho$, $\alpha$, and $\lambda$ coming from the interactions of the two scalar fields, we have introduced two parameters, $\beta$ and $\gamma$, to describe the non-minimal couplings of the scalar fields to gravity in the potential terms of ${\cal L}_V$. In total, there are 9 free parameters. To guarantee that the potential of the system has a bound below, here we assume that $\rho >0$ and $\lambda >0$ while $\alpha$, $\beta$, and $\gamma$ can be either positive or negative.

While the fermionic part can be put as 
\begin{equation}
     \begin{aligned}[t]
 {\cal L}_F =&  {\cal L}_{Fk} + {\cal L}_Y \,,\\
 {\cal L}_{Fk} = &  \sum^{f = q,\,l}_{g,\,g' = 1,\,2,\,3} (\bar{\Psi}_{L}^{g f} \hat{\slashed{D}}\Psi_{L}^{gf} +  \bar{\Psi}_{1 R}^{g q} \hat{\slashed{D}}\Psi_{1 R}^{g q} +  \bar{\Psi}_{2 R}^{g f} \hat{\slashed{D}}\Psi_{2 R}^{gf}) \\
 & + \cdots \\
- {\cal L}_Y =& 
    \sum^{f = q,\,l}_{g,\,g' = 1,\,2,\,3} (Y_{g g'}^{q} \bar{\Psi}_{L}^{' g q} \tilde{\phi} \Psi_{1R}^{g' q}  + Y_{gg'}^f \bar{\Psi}_L^{gf} \phi \Psi_{2 R}^{g' f }) \\
    & + \dots + h.c.,
      \end{aligned}
        \label{L_Orig2}
\end{equation} 
where the fermion fields $\Psi_{L}^{g f}$ for the left-handed $SU(2)$ doublet and $\Psi_{i R}^{g f}$ for the right-handed $SU(2)$ singlet are introduced. The index $i = 1,2$ corresponds to the upper and lower components of the doublet, where $f$ denotes the fermion type (such as quarks $q$ or leptons $l$), and $g$ (or $g'$) labels the fermion generations. Here only the fermionic part of the SM is presented.
New physics can be introduced by including new fermions and their new interactions with the dilaton $\Phi$, as denoted by $\cdots$ in both fermionic kinetic and Yukawa terms. 

The squared Weyl tensor $\hat C_{\mu\nu\rho\sigma }^2$, the Weyl scalar curvature $\hat R$ and the derivative $\hat \nabla_\mu$ of  the scalar  can be further expressed as the quantities of Riemannian geometry with the Weyl vector $\omega_\mu$.
\begin{equation}
\begin{aligned}
     \hat C_{\mu\nu\rho\sigma}^2 &= C_{\mu\nu\rho\sigma}^2 + 6 \,  F_{\mu\nu}F^{\mu \nu}\,,\\
 \hat R &= R - 6  (\omega_\mu \omega^\mu + \nabla_\mu \omega^\mu)\, ,\\
 \hat \nabla_\mu \phi & = \nabla_\mu \phi + d_\phi \omega_\mu \phi = \nabla_\mu \phi - \omega_\mu \phi\,, 
  \label{W2R}
\end{aligned}
\end{equation}
where $\nabla_\mu$ is Riemannian derivative and the Weyl weight $d_\phi=-1$ is given by the transformation in Eq.~\eqref{eq: ScaleTransformation}. $C_{\mu\nu\rho\sigma}$ is the traceless part of Riemannian curvature tensor $R_{\mu \nu \rho \sigma}$, known as Riemannian Weyl-tensor. Recci scalar is defined as $R= g^{\mu \nu} R_{\mu\nu}$, the second-rank Recci tensor is defined as $R_{\mu\nu} = R^{\alpha}_{\mu \alpha \nu}$ and the four-rank Reccis tensor is defined as $R^{\alpha}_{\beta \mu \nu} = \partial_\mu \Gamma^{\alpha}_{\beta \nu} - \partial_\nu \Gamma^{\alpha}_{\beta \mu} + \Gamma^{\alpha}_{\gamma \mu } \Gamma^{\gamma}_{\beta \nu} - \Gamma^{\alpha}_{\gamma \nu } \Gamma^{\gamma}_{\beta \mu}$, which are connected with the Eular-Guass-Bonnet term \cite{Chern1944,Chern1945} showing the completeness of Eq.~\eqref{L_Orig1}.  All of them are dependent upon the metric $g_{\mu\nu}$ which satisfies the metric compatibility and the Levi-Civita connection $\Gamma^{\rho}_{\mu\nu}$ which is defined as $\Gamma^\rho_{\mu\nu} = \frac{1}{2} g^{\rho\alpha}(\partial_{\mu} g_{\nu\alpha} + \partial_{\nu} g_{\mu\alpha} - \partial_{\alpha} g_{\mu\nu})$. 

The antisymmetric tensor $\hat F_{\mu\nu}$ is related to the Weyl vector field $\omega_\mu$, and is defined as $\hat F_{\mu\nu} = F_{\mu \nu}$ because of torsionless. Thus, the tensor $F_{\mu\nu}$ can be simplified as $F_{\mu\nu}= \partial_\mu \omega_\nu - \partial_\nu \omega_\mu$. And $W_\mu^a$ denotes the gauge fields (weak and hyper-charged vector fields) of the SM.

The Lagrangian is constructed based on local gauge symmetries, encompassing not only the SM but also the scale transformation in the following equation \eqref{eq: ScaleTransformation}, which is natural to consider within the Weyl geometry.
\begin{equation}
\begin{aligned}
g_{\mu\nu} &\to \Omega^2 g_{\mu\nu}\,, \quad & g^{\mu\nu} &\to \Omega^{-2} g^{\mu\nu}\,, \\
\hat{R} &\to \Omega^{-2} \hat{R}\,, \quad & \hat{C}_{\mu\nu\rho\sigma} &\to \Omega^{-2} \hat{C}_{\mu\nu\rho\sigma}\,, \\
\Phi &\to \Omega^{-1} \Phi\,, \quad  & \varphi  & \to \Omega^{-1} \varphi\,,\\
\Psi &\to \Omega^{-3/2} \Psi\,, \quad &
\omega_\mu &\to \omega_\mu - \partial_\mu \ln\Omega\,,\\
\end{aligned}
\label{eq: ScaleTransformation}
\end{equation}
Such a local gauged Weyl conformal transformation is also called as $D(1)$ symmetry, which is an Abelian type symmetry as demonstrated from this definition. It is noteworthy that the local gauge fields of the SM $W_{\mu}^a$ are invariant under such a transformation. 

This local gauged conformal symmetry must be broken at some energy scale. In literature, there are three methods of conformal symmetry breaking. 
\begin{itemize}
    \item  The first one is Coleman-Weinberg mechanism \cite{Coleman:1973jx}. Although the potential remains invariant under conformal transformations at tree level, the inclusion of quantum corrections to the Lagrangian breaks this conformal symmetry through dimensional transmutation.
    \item The second method is to introduce some terms with dimensional parameters, which explicitly break the conformal symmetry \cite{Rattazzi:2000hs}.
    \item The third one is Stueckelberg mechanism, where without knowing how the conformal symmetry is broken dynamically, a pseudo-Nambu-Goldstone particle can be introduced. The Weyl vector field becomes massive after eating the this Goldstone field \cite{PhysRevLett.61.2182,Ruegg:2003ps,ghilencea_stueckelberg_2020}. 
\end{itemize}

In this work, we adopt the third method to elucidate the spontaneous breaking of conformal symmetry. This breaking generates a non-vanishing mass term for the scalar field, which provides a seed for the spontaneous symmetry breaking of EW gauge symmetry in the SM. 

Notice that the scalar-tensor Lagrangian given in Eq.~\eqref{L_Orig1} is quadratic, which is a good candidate for quantum gravity. However, in order to return back to  the GR, we can define a Brans-Dicke (BD) field $\Theta$ to linearize the Lagrangian since all quadratic theory, or more generally, for modified f(R) theory, can be linearized as a BD type theory \cite{Capozziello:2011et}:
\begin{equation}
\Theta^2 :=\chi'^2_D \Phi^2 + \chi'^2_H H^2 \,,
\label{eq:dilaton_Higgs}
\end{equation}
where $ H^2 = 2 \phi^\dagger \phi$ denotes the modulus of the complex doublet.  When the dilaton and doublet acquire vacuum expectation values (VEVs), the Brans-Dicke field also develops a VEV $f$ causing the breaking of the local conformal symmetry.

Apparently, there could exist three different energy scales in the model. One scale is $f_D$, i.e. the vacuum expectation value (VEV) of $\Phi$. The other is $f_H$, the VEV of $\phi$. The third scale is $f$, which can be related to $f_D$ and $f_H$ via the following relation 
\begin{equation}
f^2 = \chi'^2_D \,\, f_D^2 + \chi'^2_H \,\, f_H^2
\label{fvev}
\end{equation}
Due to the nonmiminal coupling, $f^2$ could be related to the scale of gravity, i.e. $\Lambda_{\textrm{Planck}}$, though it is not necessarily. Obviously, either $f_D$ or $f_H$ or both, can trigger the conformal symmetry breaking. 

Meanwhile, it should be pointed out that the parameters $\chi'^2_D$ and $\chi'^2_H$ are not necessarily $O(1)$, i.e. they might be large or huge numbers, saying $10^{36}$ if we want to generate the Planck energy scale from the electroweak scale, which is taken as 246 GeV, as demonstrated in \cite{PhysRevLett.61.2182}. When $f$ is assumed to be around $ \sim \Lambda_{\textrm{Planck}}$, the model can be an inflaton model as investigated by Refs.\cite{ghilencea_standard_2021}. 

After the conformal symmetry breaking, we can parametrize $\Theta^2$ as 
\begin{equation}
\Theta^2 := f^2 \Sigma^2\, ,
\end{equation}
with $\Sigma^2 =\Theta^2/\langle\Theta^2\rangle =  \Theta^2/f^2$, which is a dimensionless field. From Eq.~\eqref{L_Orig1}, parameters $\chi'^2_D$ ($\chi'^2_H$) can be found which  include $\beta$ ($\gamma$) and also $\xi^2$ due to the linearization of the quadratic effect. In literature, the field $\Sigma$ can be parametrized in a nonlinear form as  
\begin{equation}
\Sigma:= \exp\left [ \frac{d}{f} \right ]\, ,
\end{equation}
{In this parametrization, $\Sigma$ plays the role of Goldstone boson and it can be eaten by the Weyl vector boson which becomes massive.}

 There is one point worthy of stress. A conformal coupling constant $\frac{D-2}{4(D-1)}$ is typically required for a $D$-dimensional non-minimal coupling scalar theory \cite{Callan:1970ze}. Thus, it seems that there should be a constraint on the free parameters of our model. However, it turns out this is not a constraint. Before the local scaling symmetry breaking, due to Weyl’s geometry, the local conformal symmetry is preserved and it is observed that the factor \textbf{6} in the second equation of \eqref{W2R} aligns with the conformal coupling constant. Therefore, there is no constraint on $\beta$ and $\gamma$. After the local conformal symmetry being broken, typically the nonminimal coupling between BD scalar and R can be parametrized as $\omega$, which is assumed to be a free parameter, as a common practice for BD theory \cite{faraoniCosmologyScalartensorGravity2004}. 

By using the following gauge fixing conditions (unitary gauge)
\begin{equation}
{\bar g}_{\mu\nu} = \Sigma^2 g_{\mu\nu}\,, \quad
{\bar \omega}_\mu = \omega_\mu - \partial_\mu \ln\Sigma\,,
\label{eq: gaugeTransformation}
\end{equation}
where $\bar{g}_{\mu\nu}$ and ${\bar \omega}_\mu$ are physical fields, we can eliminate the Goldstone of conformal symmetry breaking from the Lagrangian. 

 Here it is helpful to count the degrees of freedom of the model before and after $\Theta$ develops VEV. Before the symmetry breaking, there are four degrees of freedom in $\phi$ and one in $\Phi$, i.e. $\Theta$ includes a combination of 5 degrees of freedom. Below we can consider two cases where the symmetry breaking is driven dominantly either by the dilaton $\Phi$ or by the doublet $\phi$. 
\begin{itemize}
    \item In the first case where the symmetry breaking is dominantly driven by the dilaton $\Phi$, $f^2 \approx \chi^2_D f_D^2$, and the Goldstone $d$ is from $\Phi$. After taking unitary gauge, at the low energy region, only doublet $\phi$ is left and a mass term is generated. After $\phi$ develops VEV, the electroweak symmetry $SU(2)_L\times U_Y(1)$ is broken, and after taking unitary gauge for weak gauge bosons, only one degree of freedom is left, which corresponds to $H_{125}$.
    \item In the second case where the symmetry breaking is dominantly driven by the doublet $\phi$, $f^2 \approx \chi^2_H f_H^2$, and the Goldstone $d$ is from the doublet $\phi$. After taking unitary gauge for Weyl vector and weak vector gauge fields, at the low energy region, only the dilaton $\Phi$ is left. After $\Phi$ develops VEV, the $D(1)$ symmetry is broken, only one degree of freedom is left, which is the $H_{125}$. If this case is the reality, it is an issue how this singlet couples to the SM fermions. Thus new physics must be needed in order to cure this issue.
\end{itemize} 

\section{Two Scenarios of the model}
After BD filed $\Theta$ develops its VEV, there exist two possible parametrizations. One is  
\begin{equation}
    \Theta^2 = f^2 \Sigma^2 = f^2 \Sigma^2  \sin^2\theta  + f^2 \Sigma^2 \cos^2\theta\,,\label{case1}
\end{equation}
and the other is
\begin{equation}
    \Theta^2 = f^2 \Sigma^2 = f^2 \Sigma^2  \cosh^2\theta  - f^2 \Sigma^2 \sinh^2\theta\,, \label{case1}
\end{equation}
where $\theta$ is an angle which represents the degree of freedom left in the low energy region. Which form of $\Theta^2$ should be taken depends upon the relative signs of $\chi^2_D$ and $\chi^2_H$.

Below we systematically analyse the scenarios in term of the relative signs of $\chi'^2_D$ and $\chi'^2_H$. \\
\begin{table}[ht]
    \centering
    \begin{tabular}{|c|c|c|} \hline 
         Scenarios &  $\chi'^2_D>0$&  $\chi'^2_D < 0$\\ \hline
         $ \chi'^2_H >0 $& TSS1  & HSS2\\ \hline
         $\chi'^2_H <0$ & HSS1 & TSS2\\ \hline
         \end{tabular}
    \caption{The definitions of scenarios are tabulated. }
    \label{Tab:scenarios}
\end{table}
\indent
Although the specific form of $\chi'_{H,D}$ depends on the linearization, Eq.~\eqref{L_Orig1} gives $\chi'^2_D = \beta + k_D$ and $\chi'^2_H = \gamma + k_H$, where $k_{D/H}$ represent functions designed to linearize the quadratic curvature term.  Since the signs of $\beta$ and $\gamma$ are not specified, the signs of $\chi'^2_D$ and $\chi'^2_H$ remain undetermined, resulting in four possible scenarios. 1) the first trigonometric scalar scenario (TSS1) with $\chi'^2_D>0$ and $\chi'^2_H > 0$, 2) the second trigonometric scalar scenario (TSS2) with $\chi'^2_D < 0$ and $\chi'^2_H < 0$, 3) the frist hyperbolic scalar scenario (HSS1) with $\chi'^2_D>0$ and $\chi'^2_H<0$, and 4) the second hyperbolic scalar scenario (HSS2) with $\chi'^2_D < 0$ and $\chi'^2_H > 0$, as tabulated in Table~\ref{Tab:scenarios}.
\begin{table}[!h]
    \centering
    \begin{tabular}{|c|c|} \hline 
         $\kappa$ &  TSS \\ \hline \hline
         $\kappa_v $& $\frac{U_t}{v} \arcsin[s]/s$ \\ \hline \hline
 $\kappa_{m_W}$& $\kappa_v\sqrt{1-\chi s^2}$\\ 
 \hline 
 $\kappa_{m_h}$& $\frac{T_t}{\mu}\sqrt{1-s^2}$\\\hline
        $\kappa_{V}$  & $\kappa_v[1- \chi \,s(\sqrt{1-s^2}\arcsin[s]+s)]$\\ \hline  \hline
         $\kappa_f$&   $\frac{1}{\kappa_{m_W}} \frac{\arcsin s}{s} \sqrt{(1-s^2)(1-\chi s^2)}$
  \\\hline \hline
       $\kappa_{2V}$  & $1-\chi \, [\arcsin^2[s](1-2s^2)+4 \arcsin[s] s \sqrt{1-s^2}+s^2]$\\ \hline
       $\kappa_{3h}$& $\frac{F_t U_t}{\lambda_{SM} v} (1-2s^2)\sqrt{1-s^2}$\\ \hline
       $\kappa_{4h}$ & $\frac{F_t}{\lambda_{SM}}(1-\frac{28}{3}s^2(1-s^2))$\\ \hline
    \end{tabular}
    \caption{The $\kappa$ parameters in TSS are listed. }
    \label{Tab:kappa_Tri}
\end{table}
Owing to the reason that the TSS2 can not produce the correct sign in the GR, we will not consider it here. Thus, we can label TSS1 as TSS for simplicity. Also it should be noticed that the parametrization is democratic for either the real scalar and the complex doublet, which means the similarity for HSS1 and HSS2. Therefore, we can just consider HSS1 without losing generality and define it as HSS in the discussion below.\\
\indent
In the TSS, we can parametrize doublet and dilaton scalar fields by $\chi^{2}_H =\chi'^2_H>0$ and $\chi_D^{2} = \chi'^2_D>0$ into the following form
\begin{equation}
    H^2 = \frac{f^2}{\chi_H^2}  \Sigma^2 \sin^2\theta\,,\quad \Phi^2=\frac{f^2}{\chi_D^2}  \Sigma^2 \cos^2\theta\,. \label{case1}
\end{equation}
Substituting complex and dilaton fields into the Lagrangian and using the unitary gauge given in Eq.~\eqref{eq: gaugeTransformation},
we can arrive at a Lagrangian with only physical fields.

The Higgs potential can be computed as given below: 
\begin{widetext}
    \begin{equation}
        \begin{aligned}
            V &= \frac{f^4 \rho}{4!\chi_D^4}-\frac{T_t^2}{2}\frac{f^2}{\chi_H^2}\sin^2[\frac{\chi_H}{f}\theta']+\frac{F_t}{4}\frac{f^4}{\chi_H^4}\sin^4[\frac{\chi_H}{f}\theta']\,,\\
              & \approx 
                  \frac{f^4 \rho}{4!\chi_D^4} -\frac{F_t U_t^4}{4} +T_t^2(1-s^2) h^2  + F_t U_t \sqrt{1-s^2}(1-2s^2) h^3 -\frac{F_t}{4} \Big(1 - \frac{28s^2}{3} (1 - s^2)\Big) h^4 + \dots \, .
        \end{aligned}
    \label{Potential_Tri_Poly}
\end{equation}
\end{widetext}
where the $\cdots$ denote high order terms, like $h^6$ and $h^8$ or higher terms. In order to cast the Higgs potential into the standard form, some short-handed parameters are defined and given below 
\begin{eqnarray} 
        T_t^2 &=& \frac{f^2}{2\chi_D^2}(\frac{\rho \chi_H^2}{3\chi_D^2}-\alpha)>0 \,, \nonumber\\
       F_t &=& \lambda+\frac{\rho \chi_H^4}{6\chi_D^4}-\frac{\alpha \chi_H^2}{\chi_D^2}> 0 \,, \qquad U_t \coloneqq \sqrt{\frac{T_t^2}{F_t}}\,, \label{Eff_Coup_Tri}\\
    \theta' &= & \frac{f}{\chi_H} \theta \coloneqq f_{EW}+h \,,\qquad f_{EW} = \frac{f}{\chi_H}\arcsin[\frac{\chi_H}{f}U_t]\,. \nonumber
\end{eqnarray}
\indent
The full form of the Higgs potential of TSS is shown as the first line, which is periodic due to the periodicity of the trigonometric functions. While the form in the second line is obtained by using Taylor expansion. It should be pointed out that the expansion is based on $\sin[\frac{\chi_H}{f}h]$, not on $\sin[\frac{\chi_H}{f}\theta']$ instead. Here $h$ denotes $H_{125}$ and the parameter $U_t$ is the VEV of the trigonometric scenario of the electroweak scalar $\sin[\frac{\chi_H}{f}\theta]$ in our model.\\
\indent
The BSM deviations of Higgs couplings are described using the $\kappa$-framework, and by analogy, we also define the deviations in the VEV\footnote{Notice that $\kappa_v =  \frac{f_{EW}}{v}$ does not need to equal 1 because $f_{EW}$ is defined before canonicalization.} and the masses of the W boson and the Higgs boson which are listed with all relevant $\kappa$ parameters in Table~\ref{Tab:kappa_Tri}.

It should be pointed out that these $\kappa$ parameters are characterized by two parameters, i.e. $\chi$ and $s$, which are defined as given below
\begin{equation}
    \begin{aligned}
        \chi &\coloneqq 1-\frac{\chi_H^2}{\chi_D^2} < 1\,,\\
        s & \coloneqq  = \frac{\chi_H U_t}{f} =\sin[\frac{\chi_H}{f}f_{EW}]> 0\,,
    \end{aligned}
    \label{eq:TSSparameters}
\end{equation}
where the parameter $\chi$ measures the discrepancy of scalars, it is 1 when $\chi_H^2\to0$ and zero when $\chi_H^2=\chi_D^2$. 
While the parameter $s$ measures the scale symmetry breaking caused by the doublet $\phi$, and it is 0 when $\chi_H\to 0$ and $1$ when $\chi_H U_t \to f$. It is symmetric for $s>0$ and $s< 0$ due to a $Z_2$ symmetry of Higgs potential. For the sake of simplicity, we will focus on the case where $s>0$.

For the HSS, scalar fields can be parametrized using $\chi_H^2 = -\chi'^2_H >0$ and $\chi_D^2=\chi'^2_D>0$ as follows
\begin{equation}
    H^2 = \frac{f^2}{\chi_H^2}  \Sigma^2 \sinh^2\theta\,,\quad \Phi^2=\frac{f^2}{\chi_D^2}  \Sigma^2 \cosh^2\theta\,.
    \label{case2}
\end{equation}
\\
Substituting these parametrization and using the unitarity gauge to the Lagrangian, the corresponding Higgs potential can be obtained:
\begin{widetext}
    \begin{equation}
        \begin{aligned}
              V&= \frac{f^4 \rho}{4!\chi_D^4}-\frac{T_h^2}{2}\frac{f^2}{\chi_H^2}\sinh^2[\frac{\chi_H}{f}\theta']+\frac{F_h}{4}\frac{f^4}{\chi_H^4}\sinh^4[\frac{\chi_H}{f}\theta'] \,,\\
              &\approx \frac{f^4 \rho}{4!\chi_D^4} -\frac{F_h U_h^4}{4} +T_h^2(1+s^2) h^2  + F_h U_h \sqrt{1+s^2}(1+2s^2) h^3 + \frac{F_h}{4}\Big(1 + \frac{28 s^2}{3} (1 + s^2)\Big) h^4 + \dots \,,
        \end{aligned}
    \label{Potential_Hyp_Poly}
\end{equation}
\end{widetext}
with the following parameters being defined
\begin{eqnarray} 
     T_h^2 &=& - \frac{f^2}{2 \chi_D^2}(\alpha+\frac{\rho \chi_H^2}{3 \chi_D^2})>0  \,, \nonumber\\
       F_h &=&  \lambda + \alpha \frac{\chi_H^2}{\chi_D^2}+\frac{\rho}{6}\frac{\chi_H^4}{\chi_D^4} >0 \,, \qquad  U_h \coloneqq \sqrt{\frac{T_h^2}{F_h}}\,,  \label{Eff_Coup_hyp}\\
       \theta' &= & \frac{f}{\chi_H} \theta \coloneqq f_{EW} + h\, , \qquad f_{EW} = \frac{f}{\chi_H}\arcsinh[\frac{\chi_H}{f}U_h]\,. \nonumber
\end{eqnarray}
\indent
Obviously, the potential of HSS possesses no periodicity, in contrast to that of the TSS. The potential can be expanded by using the Taylor expansion in term of $\sinh[\frac{\chi_H}{f}h]$ rather than $\sinh[\frac{\chi_H}{f} \theta']$, which is given in the second line. One should realize that the difference for these two scenarios lies in the difference of trigonometric and hyperbolic functions.

From the Lagrangian, we extract the $\kappa$ values and present them for the HSS in Table~\ref{Tab:kappa_Hyp} which are dependent upon two parameters $\chi$ and $s$ being defined below:
\begin{equation}
    \begin{aligned}
        \chi &\coloneqq 1+\frac{\chi_H^2}{\chi_D^2}> 1\,,\\
        s & \coloneqq \frac{\chi_H U_h}{f} = \sinh [\frac{\chi_H}{f}f_{EW}] > 0\,.
    \end{aligned}
    \label{eq:chi_h}
\end{equation}
\begin{table}[h]
    \centering
    \begin{tabular}{|c|c|} \hline 
         $\kappa$ &  HSS  \\ \hline \hline
         $\kappa_v $& $\frac{U_h}{v} \arcsinh[s]/s$\\ \hline \hline
 $\kappa_{m_W}$& $\kappa_v\sqrt{1+\chi \,  s^2}$\\ \hline 
 $\kappa_{m_h}$& $\frac{T_h}{\mu}\sqrt{1+s^2}$\\\hline
        $\kappa_{V}$  & $\kappa_v[ 1+\chi\,  s(\sqrt{1+s^2}\arcsinh[s]+s) ]$\\ \hline \hline
        $\kappa_f$&  $\frac{1}{\kappa_{m_W}} \frac{\arcsinh s}{s} \sqrt{(1+s^2)(1+\chi s^2)}$\\\hline\hline
       $\kappa_{2V}$  & $1+\chi\,  [ \arcsinh^2[s](1+2s^2)+4 \arcsinh[s] s \sqrt{1+s^2}+s^2 ] $\\ \hline
       $\kappa_{3h}$& $\frac{F_h U_h}{\lambda_{SM} v} (1+2s^2)\sqrt{1+s^2}$\\ \hline
       $\kappa_{4h}$ & $\frac{F_h}{\lambda_{SM}}[ 1+\frac{28}{3}s^2(1+s^2) ] $\\ \hline
    \end{tabular}
    \caption{The $\kappa$ parameters in the HSS are tabulated.}
    \label{Tab:kappa_Hyp}
\end{table}
\indent
\section{Numerical Analysis on the dilaton dominance}
\begin{figure}
    \centering
    \includegraphics[width=1\linewidth]{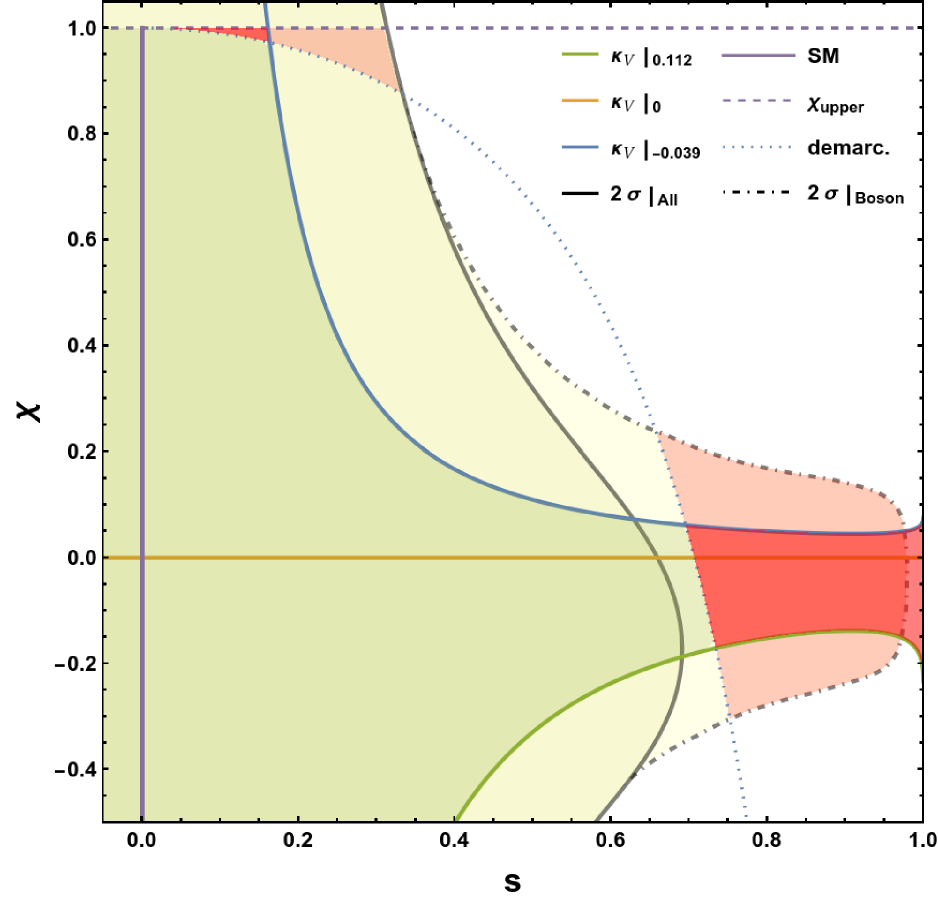}
    \caption{Allowed parameter space by experiments for TSS in $\chi$ and $s$ plane is shown.}
    \label{fig:PossibleSolutionsTri}
\end{figure}

\begin{figure}
    \centering
    \includegraphics[width=1\linewidth]{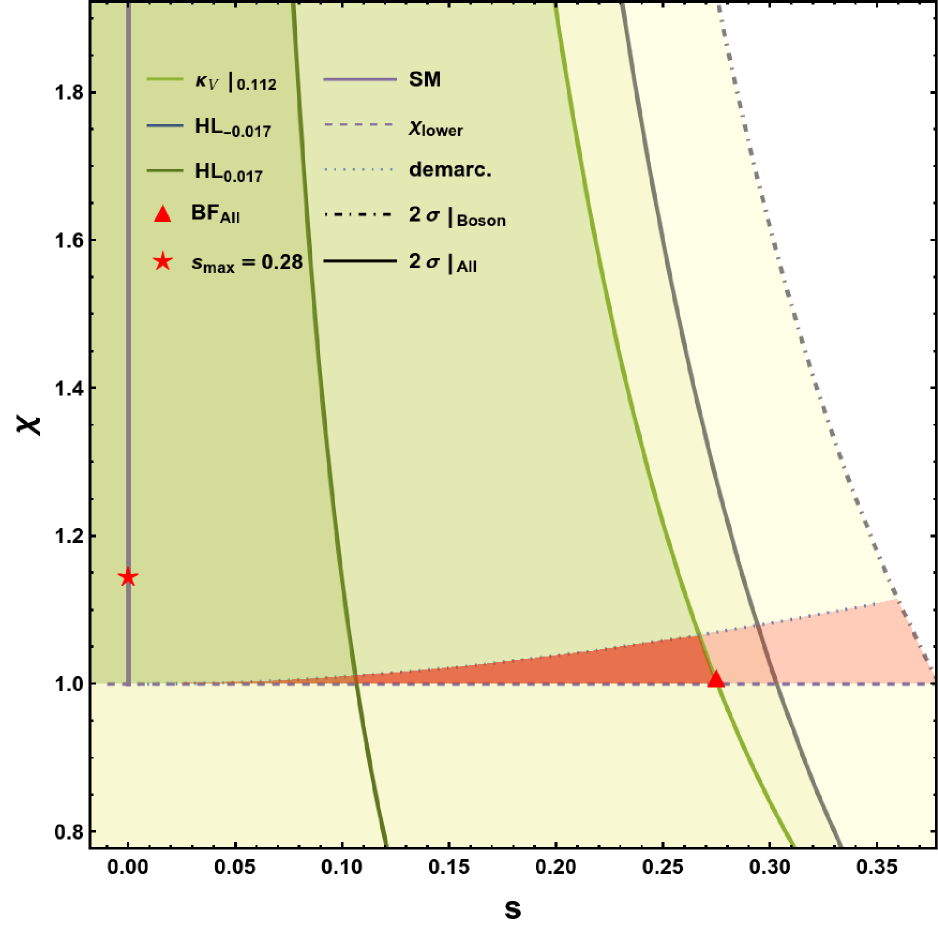}
    \caption{Allowed parameter space in the HSS is shown.}
    \label{fig:PossibleSolutionsHyp}
\end{figure}

We adopt two analysis methods: one is a global fitting (GF) method, the other is specific solution (SS) method. Below we examine these two scenarios by using the LHC experimental measurements presented in Table~\ref{tab:lhc}. After taking into account the requirement of Eq.~\eqref{eq:TSSparameters}/\eqref{eq:chi_h}, the allowed parameter space in the $\chi-s$ plane determined by these two methods can be found for these two scenarios in Fig. \ref{fig:PossibleSolutionsTri} and Fig. \ref{fig:PossibleSolutionsHyp}. \\
\indent
 To examine the effects of Yukawa couplings, we present the global fitting results with all data (including $\kappa_f$ measurements) given in Table~\ref{tab:lhc} and the fitting results without $\kappa_f$ measurements. We consider a fit with 4 free parameters, i.e. $s$, $\chi$, $T$, and $F$, by using the relations in Tabs.~\ref{Tab:kappa_Tri} and \ref{Tab:kappa_Hyp}. When the results are projected on the $s-\chi$ plane, for the TSS scenario, the best-fit values from all measurements (without $\kappa_f$ data) are given as 
$s=0.07\, (0.04)$ and $\chi=-7.46\,(-10.65)$, with the $\chi^2_{\text{min}} = 1.93\,(0.65)$; While for the HSS scenario, the best-fit values are given as $s=4.16\,(3.50) \times 10^{-5}$ and $\chi=1.14 \, (1.15)$, with $\chi_{\text{min}}^2 = 1.94\,(0.84)$ indicated by the five-angle star in the figure. All fits are considered good since their p-values exceed 50\%. The globel fitting results indicate that the model can accommodate the experimental data quite well.\\
\indent
From our results, it is also clear that in essential, our global fitting demonstrates that the SM is favoured by the experimental data, which corresponds to the limit $s\rightarrow 0$ and is represented by the purple line in each of these two figures.\\
\indent
The 2$\sigma$ contours of the fits are shown, with the color shaded region bounded by a dot-dashed line corresponding to the results using only bosonic processes without the $\kappa_f$ data, whereas the color shaded region bounded by a thick solid line corresponds to the results including all LHC measurements.\\
\indent
According to the definitions in Eqs.~\eqref{case1} and \eqref{case2}, the dividing line of doublet-dominant or dilaton-dominant is determined by Eq.~\eqref{criterion}
\begin{equation}
    \begin{aligned}
      \frac{\Phi^2}{H^2}|_{TSS}&\to \frac{\chi_H^2}{\chi_D^2}\frac{\cos [\frac{\chi_H}{f}\theta]}{\sin[\frac{\chi_H}{f}\theta]}|_{vev} = (1-\chi)\frac{1-s^2}{s^2}=1,\\
      \frac{\Phi^2}{H^2}|_{HSS}&\to \frac{\chi_H^2}{\chi_D^2}\frac{\cosh [\frac{\chi_H}{f}\theta]}{\sinh[\frac{\chi_H}{f}\theta]}|_{vev} = (\chi-1)\frac{1+s^2}{s^2}=1,
    \end{aligned}
    \label{criterion}
\end{equation}
and the parameter regions where the ratio is smaller (larger) than 1 is defined as dilaton (doublet) dominance. These two parameter regions is separated by the dotted line in these figures, while the doublet-dominant region is shaded in blue and the dilaton-dominant region is shaded in red. 
\begin{table}
    \centering
    \begin{tabular}{|c|c|}\hline
       $\kappa_{m_W}$  & $1.00014^{+0.00021}_{-0.00021}$ (99.98\%CL)\cite{ATLAS:2024erm}\\\hline
       $\kappa_{m_h}$& $1.00176^{+0.001118}_{-0.001118}$ (99.89\%CL)\cite{ATLAS:2023owm} \\\hline
       $\kappa_{V}$  &  $1.035^{+0.077}_{-0.074}$ (95\%CL)\cite{aad_detailed_2022} \\\hline \hline
       $\kappa_{t}$  & $0.99^{+0.09}_{-0.08}$ (90\%CL)\cite{ATLAS-CONF-2025-006}\\\hline
       $\kappa_{b}$  & $0.90^{+0.12}_{-0.11}$
       (90\%CL)\cite{ATLAS-CONF-2025-006} \\\hline
       $\kappa_{\tau}$  & $0.95^{+0.10}_{-0.08}$ (90\%CL)\cite{ATLAS-CONF-2025-006} \\\hline
       $\kappa_{\mu}$  & $1.05^{+0.25}_{-0.31}$ (90\%CL)\cite{ATLAS-CONF-2025-006} \\ \hline \hline
       $\kappa_{2V}$  & $0.92^{+0.58}_{-1.52}$ (95\%CL)\cite{ATLAS:2024ish} \\\hline
       $\kappa_{3h}$  & $4.3^{+2.9}_{-5.5}$ (95\%CL)\cite{ATLAS:2024ish} \\\hline
    \end{tabular}
    \caption{The values of $\kappa$ from LHC measurements are provided.}
    \label{tab:lhc}
\end{table}

Apart from the GF method, we also perform a detailed analysis by using the SS method, in which we use the mass of the W boson \cite{atlascollaboration2024measurementwbosonmasswidth}, the mass of the Higgs boson \cite{atlascollaboration2024characterisingHiggsbosonatlas}, and the coupling $\kappa_V$ \cite{aad_detailed_2022} as model inputs to to solve out the border of allowed  parameter space. In practice, we fix $\kappa_{m_W} = 1$, and $\kappa_V = 1 + \delta$ (95\% CL) is taken from the experimental results reported in \cite{aad_detailed_2022}:
\begin{equation}
        \kappa_V =  \,1.035^{+0.077}_{-0.074}\, = \, 1^{+0.112}_{-0.039}\,.
        \label{deltakappav}
\end{equation} 
 The projected precision of HL-LHC $\kappa_V = 1\pm 0.017$ (68$\%$ CL) \cite{Dainese:2019rgk} are also illustrated as darker lines and shaded region in Fig. \ref{fig:PossibleSolutionsHyp}. 

\begin{figure}
    \centering
    \includegraphics[width=1\linewidth]{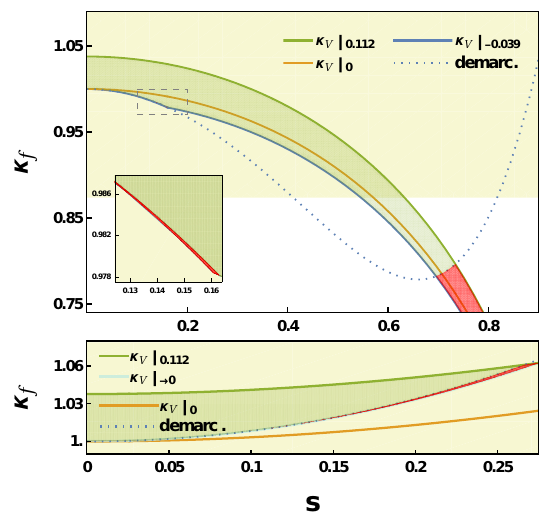}
    \caption{The constraints on $\kappa_{f}$ in the TSS(upper)/HSS(lower) scenarios are shown.}
    \label{fig:fermions}
\end{figure}

In the SS method, the measurements of Yukawa couplings require $\kappa_f=0.979_{-0.105}^{+0.114}\, (95\% \,\text{CL})$ \cite{ATLAS-CONF-2025-006} which constrain the model parameters, like $s$, as shown in Fig.~\ref{fig:fermions} by the yellow shaded region in the $\kappa_f -s$ plane. The upper and lower panels of this figure correspond to the TSS and HSS scenarios, respectively. 
It is obtained by assuming $\kappa_{m_W}=1$, with the fermion masses fixed to their SM values, and the red shaded region denotes the parameter space of dilaton dominance (the legend is the same as in Fig.~\ref{fig:PossibleSolutionsTri} and Fig.~\ref{fig:PossibleSolutionsHyp}).

As illustrated in Fig.~\ref{fig:fermions}, the dilaton-dominant interpretation still exists, although a larger fraction of parameter space favors the doublet-dominant interpretation of $H_{125}$. Such a conclusion also hold when all Higgs data are taken into account and is consistent with the global fitting results given in Figs.~\ref{fig:PossibleSolutionsTri} and Fig.~\ref{fig:PossibleSolutionsHyp}.

 In the HSS scenario, the entire dilaton-dominant region lies within the 95$\%$ CL of the experimental constraint on $\kappa_f$, as shown in Fig.~\ref{fig:PossibleSolutionsHyp} and the lower panel plot of Fig.~\ref{fig:fermions}. In contrast, in the TSS scenario, there are two separate regions for the dilaton dominance, one is outside the 95$\%$ CL for $s > 0.6$, the other is for $s < 0.2$, as illustrated in the zoomed inset in the upper panel plot of Fig.~\ref{fig:fermions} and Figs.~\ref{fig:PossibleSolutionsTri}. The zoomed inset in the upper panel plot of Fig.~\ref{fig:fermions} highlights the viable region in the $s \to 0$, $\chi \to 1$ regime shown in Fig.~\ref{fig:PossibleSolutionsTri}, corresponding to the limit $\chi_H \to 0$ where $f^2 \approx \chi_H^2 f_H^2$ given $f_H \gg f_D$. This parameter region is narrow but is still allowed, which could be ruled out only if the HL-LHC measurement would exclude the case $\kappa_V < 1$.

 After examining the effects of Yukawa couplings, below we also examine the prediction of Higgs-weak boson couplings and Higgs self couplings of the model in the SS method.  When we fix $\kappa_{m_h} =1$, the predicted Higgs trilinear couplings can be determined as 
\begin{equation}
    \begin{aligned}
        \kappa_{3h} = &\, \frac{F U}{\lambda_{SM} v} (1-2s^2)\sqrt{1-s^2} \, ,\\
        \xrightarrow{\kappa_{m_h} = 1} &\, \frac{\mu^{2}}{\lambda_{SM} v^2}\frac{\arcsin[s]}{s \sqrt{1-s^2}}(1-2s^2)\sqrt{1- \chi s^2}\,.
    \end{aligned}
\end{equation}

In particular, for TSS, we focus on the analysis of the solution $\chi=0$, since it can lead an interesting relation $\chi_D^2 = \chi_H^2$. Consequently, the couplings between the Weyl vector and Higgs are decoupled, and $\kappa_{2V} = 1$ in TSS. More discussion of the couplings to fermions can be found in Ref.~\cite{ghilencea_standard_2021}.

\begin{figure}
    \centering
    \includegraphics[width=1\linewidth]{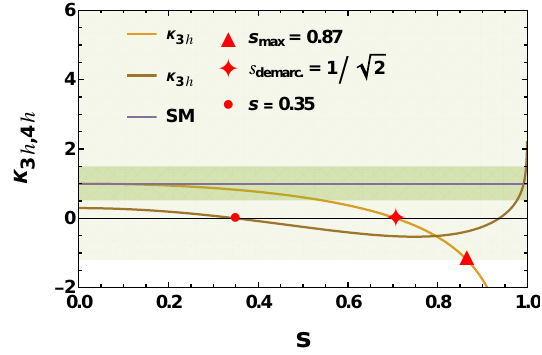}
    \caption{The varying of $\kappa_{3h/4h}$ with parameter $s$ in the TSS are shown($\chi =0$).}
    \label{fig:ExpConstraintsTri}
\end{figure}

The relationships between $\kappa_{3h}$, $\kappa_{4h}$, and $s$ are depicted by the orange and darker orange lines, respectively, in Fig.~\ref{fig:ExpConstraintsTri}. The experimental measurements of $\kappa_{3h}$ from Table~\ref{tab:lhc} are depicted as the lighter blue shaded region, constraining $s < 0.87$. Meanwhile, the projected precision of the HL-LHC \cite{Dainese:2019rgk} at 68\% CL for $0.52<\kappa_\lambda<1.50$ is also shown by the darker blue shaded region.

The four-pointed star marker separates the allowed parameter of $s$ into two parts. The first part is $0 < s< \frac{1}{\sqrt{2}} $, where the scalar $H_{125}$ corresponds to doublet dominance since $H_{125}$ is mainly from complex doublet. While the second part is $ \frac{1}{\sqrt{2}} < s \leq 0.87$, where the scalar $H_{125}$ is dilaton boson dominant from scalar singlet mainly. Notice that $s=\frac{1}{\sqrt{2}}$ is the maximum mixture sate for the doublet and dilaton, making it difficult to distinguish whether the doublet or dilaton dominates. Furthermore, it is observed that the HL-LHC run is capable to confirm/rule out the parameter space of dilaton dominance.

The dilaton-dominant region, defined by $\frac{1}{\sqrt{2}} \leq s \leq 0.87$, predicts a negative Higgs self-coupling for both three-point and four-point interactions. It is remarkable that even for the doublet-dominant region $0 \leq s \leq \frac{1}{\sqrt{2}}$, the quartic Higgs self-coupling can be either positive or negative, and a vanishing quartic coupling is also possible. This distinctive feature originates from the TSS scenario, resulting in a periodic potential.

\begin{figure}
    \centering
    \includegraphics[width=1\linewidth]{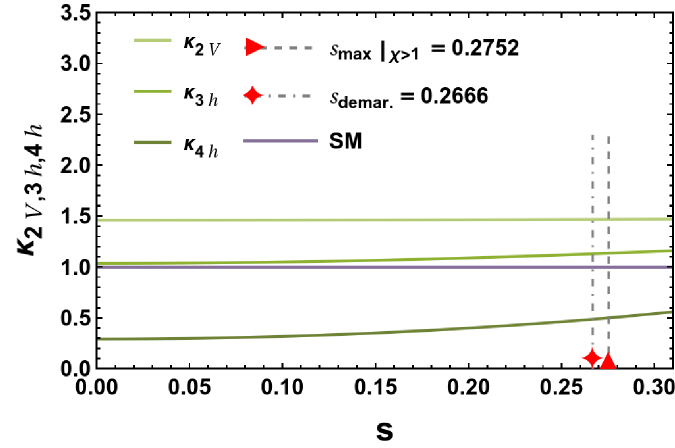}
    \caption{The constraints to $\kappa_{2V/3h/4h}$ in the HSS are shown.}
    \label{fig: ExpConstraintsHyp}
\end{figure}

For a potential of $\phi^4$ theory, a negative quartic coupling means that the vacuum is unstable. In the SM, a negative quartic coupling of Higgs potential means the potential is unbounded below \cite{Djouadi:2005gi}. However, it should be pointed out that the vacuum stability in the TSS of our model is ensured, as demonstrated by the full form of the potential given in Eq.~\eqref{Potential_Tri_Poly}. In the form of its Taylor expansion, there are higher order terms like $h^6$ and $h^8$, etc., which can guarantee the vacuum stability of our model.\\
\indent
Since $s$ is tightly constrained by the upper bound on $\kappa_V$ (as illustrated in Fig.~\ref{fig:PossibleSolutionsHyp}), our analysis emphasizes this constraint in Fig. ~\ref{fig: ExpConstraintsHyp}.
The triangular marker line indicates the maximum value of $s$, while the four-point marker line marks the demarcation between doublet- and dilaton-dominant regions along the allowed blue curve. The quantities $\kappa_{2V}$, $\kappa_{3h}$, and $\kappa_{4h}$, which depend on the upper bound of $\kappa_V$ and vary with $s$, are depicted by the light blue, blue, and darker blue lines, respectively. It is obvious that these measurements are not sensitive to the breaking of scale symmetry $s$ in HSS, which will be difficult to be measured even at the future 100 TeV colliders \cite{Chen:2015gva,Kilian:2017nio,Kilian:2018bhs}. Nonetheless, the measurements can be enhanced due to the sensitivity of $\kappa_V$, with improved precision provided by the HL-LHC, as shown in Fig.~\ref{fig:PossibleSolutionsHyp} introduced above. \\
\indent
\section{Scalars potentials and spontaneous symmetry breaking}
\begin{figure}
    \centering
    \includegraphics[width=1\linewidth]{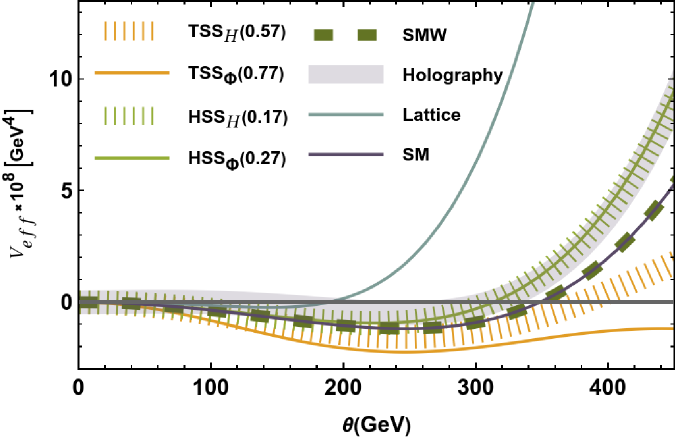}
    \caption{The shape of Higgs potentials are demonstrated.}
    \label{fig:potentials}
\end{figure}

In Table~\ref{tab:Potentials}, we also list a few typical scalar potentials in literatures which can trigger spontaneous symmetry breaking and compare the shapes of these potentials in Figure \ref{fig:potentials}.\\
\indent
In the reference \cite{ghilencea_standard_2021}, the Higgs potential in the SMW is found to have the following form
\begin{equation}
V(\sigma)=\frac{3}{2}M_p^2\Big[6 \lambda \sinh^4\frac{\sigma}{M_p \sqrt{6}} +\xi^2 (1-\xi_h \sinh^2 \frac{\sigma}{M_p \sqrt{6}})^2\Big]\,,
\label{V_SMW}
\end{equation}
where $\sigma$ is assumed to be the Higgs boson, and $M_p$ is the Planck energy scale. The parameters $\xi$ and $\xi_h$ represent two non-minimal couplings. In the limit that $\sigma \ll M_p$, the potential can be expanded as
\begin{equation}
    V(\sigma)=\frac{1}{4} (\lambda - \frac{1}{9} \xi_h \xi^2+ \frac{1}{6} \xi_h^2 \xi^2) \sigma^4 - \frac{1}{2} \xi_h \,\,\xi^2\,\, M_p^2 \,\,\sigma^2 + \cdots\,,
    \label{V_SMW}
\end{equation}
which takes the form of Higgs potential of the SM. In order to comply with the value of the Higgs mass and EW VEV, the parameters $\xi \sqrt{\xi_h}\sim 3.5 \times 10^{-17}$ and $\lambda \sim \lambda_{SM}$ are chosen as shown in the second row of Table~\ref{tab:parameters_SMW}.

\begin{table}
    \centering
    \begin{tabular}{|c|c|c|} \hline 
Framework & $\xi_h$ & $\xi$ \\ \hline 
\multirow{2}{*}{Particle Physics (set A)}  & \multicolumn{2}{c|}{$\xi \sqrt{\xi_h} \sim 3.5 \times 10^{-17}$} \\ 
                                  & \multicolumn{2}{c|}{To fix the mass of $H_{125}$} \\ \hline
\multirow{2}{*}{Cosmology (set B)}        & $10^{-3} \sim 10^{-2}$ & $10^{-9}$ \\ 
                                  & Inflation              & CMB \\ \hline
    \end{tabular}
    \caption{The typical orders of parameters in SMW\cite{ghilencea_standard_2021} to accommodate experimental data are provided.}
    \label{tab:parameters_SMW}
\end{table}

In the reference \cite{Goldberger:2007zk}, in terms of scale transformation, by introducing the spurion field into the low-energy effective theory, the general dilaton potential can be of the form
\begin{equation}
V(\chi) = \chi^4 \sum_{n=0}^{\infty} c_n (\Delta_{\cal{O}}) \left( \frac{\chi}{f}\right )^{n (\Delta_{\cal{O}} - 4) }\,,
\end{equation}
where $\chi = f + \bar{\chi}$, with $\bar{\chi}(x)$ representing the dilaton field, $f$ being the VEV of $\chi$, and $\Delta_{\cal{O}}$ denoting the dimension of the operator $\cal{O}$ that breaks conformal symmetry. And the coefficients $c_n$ depend on the dynamics of the underlying conformal field theory. In the limit $\Delta_{\cal {O}} \to 4 $, the Coleman-Weinberg correction to the potential can be of the form
\begin{equation}
V(\chi) = \frac{1}{16}\frac{m^2}{f^2}\chi^4[4 \,\,{\rm ln}\frac{\chi}{f} - 1] + \mathcal{O}(\left|\Delta_\mathcal{O} - 4\right|^2)\,.
\label{V_holo}
\end{equation}
\indent
It includes the SSB scale $f$, which is set as $f= v_{\text{SM}}$ and $m\sim m_{\text{H}_{125}}/\sqrt{2}$ GeV in order to compare and contrast with other potentials in Figure \ref{fig:potentials}.\\
\indent
In reference \cite{appelquist_dilaton_2020,appelquist_dilaton_2022}, in order to explore the dynamic of conformal symmetry breaking, an effective field theory of dilaton field is constructed by using lattice data. The effective potential of dilaton field can be cast into a form as given below
    \begin{equation}
        \begin{aligned}
V(\chi) =  & \frac{m_d^2\chi^4}{4(4-\Delta_{\mathcal{O}})f_d^2}\left[1-\frac{4}{\Delta_{\mathcal{O}}}\left(\frac{f_d}{\chi}\right)^{4-\Delta_{\mathcal{O}}}\right] \\ & -\frac{N_f m_\pi^2 f_\pi^2}{2}(\frac{\chi}{f_d})^y    \,.
\label{V_Lattice}
\end{aligned}
\end{equation}
The final term in the potential arises from light quark contributions, where $m_\pi$ and $f_\pi$ represent the mass and decay constant of the pion fields, respectively. This term is essential for the lattice method and explicitly breaks scale symmetry through the tunable mass $m_\pi$. The soft theorem suggests that $\Delta_{\mathcal{O}} \to 2$ at the infrared fixed point \cite{zwicky_qcd_2023}. Compared with Eq.~\eqref{V_holo}, these represent two extreme cases of conformal symmetry breaking caused by either a gluon condensate $\Delta_{\mathcal{O}} = \Delta_{G2} = 4$ or a fermion condensate $\Delta_{\mathcal{O}} = \Delta_{\bar{f}f} = 2$. Lattice data yield a more comprehensive result of $\Delta_{\mathcal{O}} = 3.5$ with $y =  2.06 \pm 0.05$
by using SU(3) YM theory with $N_f = 8$ Dirac fermions in the fundamental representation. The parameters $m_d$ and $m_\pi$ are fixed to determine the mass of $\text{H}_{125}$ by the definitions. All parameters for Figure \ref{fig:potentials} are presented in the Table 1 of \cite{appelquist_dilaton_2020}. 

It is noteworthy that potentials given in Eq.~\eqref{V_holo} and Eq.~\eqref{V_Lattice} can also be introduced to explicitly break the local conformal symmetry of the Lagrangian in Eq.~\eqref{L_Orig} as well, which belong to the second method. Here these two potentials serve as the template potentials to describe the dilaton-dominant cases. 
        
After choosing some parameters as given in Table~\ref{tab:Potentials}, we plot the shape of these Higgs potentials which are shown in Figure \ref{fig:potentials}. For the scenarios we discuses above, we thoroughly demonstrate each possibility. The orange curves repr, esent two TSS cases: the dotted curve is the doublet dominance with $s=0.57$ and the thick curve is dilaton dominance with $s=0.77$. Meanwhile, there are two HSS cases one for the doublet dominance with $s=0.17$ and the other one for dilaton dominance with $s=0.27$ which are shown in blue curves. The potential of SMW model is plotted by a dashed curve while the potential determined from Lattice calculation is drawn by using a darker cyan curve. A shallow purple curve is to represent the Higgs potential in the holographic framework, incorporating the Coleman-Weinberg correction. The purple curve is for the Higgs potential of the SM.

	\begin{table}[H]
		\centering
		\begin{tabular}{|c|c|c|c|} \hline
			Model & Potential &VEV (GeV)  & Parameters\\ \hline
		\multirow{2}{*}{\centering TSS}  & \multirow{2}{*}{Eq.~\eqref{Potential_Tri_Poly}}&246.3 & s = 0.57, \, $\kappa_V = 1$\\ \cline{3-4}
											 &                                                               &246.3 & s = 0.77,\, $\kappa_V = 1$\\ \hline
		\multirow{2}{*}{HSS}              &  \multirow{2}{*}{Eq.~\eqref{Potential_Hyp_Poly}}&221.64& s = 0.17, \, $\kappa_V = 1.112$\\  \cline{3-4} 
		& &223.88& s = 0.27, \, $\kappa_V = 1.112$\\ \hline \hline
        SMW \cite{ghilencea_standard_2021}&Eq.~\eqref{V_SMW}&246.3 & 
        \begin{tabular}{@{}c@{}} $\xi \sqrt{\xi_h}\sim 3.5 \times 10^{-17}$ \\ $\lambda \sim \lambda_{SM}$ \end{tabular}
        \\ \hline
		  Holography \cite{Rattazzi:2000hs, Goldberger:2007zk}   &Eq.~\eqref{V_holo}  &246.3 & \begin{tabular}{@{}c@{}} $\Delta_{\mathcal{O}} \to 4$\\$f = v_{SM}$\\ $m = m_{\text{H}_{125}}/\sqrt{2}$ \end{tabular}\\ \hline
		Lattice \cite{appelquist_dilaton_2020,appelquist_dilaton_2022} & Eq.~\eqref{V_Lattice} &144.47 & \begin{tabular}{@{}c@{}} $N_f = 8$ \\$\Delta = 3.5$ \\ $y = 2.06 \pm 0.05$ \\ $m_d = 125GeV$\\ $m_\pi = 0.2MeV$ \end{tabular} \\  \hline
		SM  & Eq.~\eqref{V_SM} &246.3 & $\lambda_{SM} = 0.13$  \\ \hline
		\end{tabular}
         \caption{Higgs potentials in a few typical models are provided and compared.}
      \label{tab:Potentials}
		\end{table}

An obvious feature is that these scalar potentials can trigger spontaneous EW symmetry breaking successfully after the conformal symmetry breaking and can accommodate Higgs data, the correspondent mass of $H_{125}$ demonstrated in the third column of Table~\ref{tab:Potentials} which is fixed at 125.11 GeV. There are a few comments in order.
\begin{itemize}
    \item 
Although the VEV of HSS is different from that of the SM, it is consistent with experimental errors. 
\item The holographic potential closely resembles the HSS type potential, although the VEVs differ between the two distinct models. The underlying reason may stem from the mathematical framework in which the scale transformation is described by a hyperbolic function. Further aspects of the physical relationship warrant additional investigation. 
\item One special exception is the Lattice type, where the VEV is 144.47 GeV. The gap between theoretical and experimental VEV is large. Since the parameters $m_\pi$ or $f_\pi$ are adjustable, the vacuum expectation value (VEV) can be accurately determined, albeit with a corresponding adjustment in the associated mass term.

\item Two key points should be highlighted, for which the potential of TSS-dilaton in Figure \ref{fig:potentials} starts to drop down after $\theta>450$ GeV. Firstly, because the TSS potential depicted is based on a vacuum expectation value (VEV) of 246.3 GeV, the expansion is less reliable for $\theta > 450$ GeV. Secondly, the full potential exhibits oscillatory behaviour, which can be attributed to the trigonometric function within Eq.~\eqref{Potential_Tri_Poly}.
\item A remarkable character of our model is that the potentials of HSS are exponentially dependent upon $h$ while those of TSS are periodic, which might have some different effects during phase transitions.
\end{itemize}

\section{Discussions and Conclusions}
In this paper, we investigate the scalar sector of a dilaton model with the local gauge conformal symmetry $D(1)$. In the model, two scenarios, TSS and HSS, are identified due to uncertainties in the linearization from quadratic gravity to general relativity.
Furthermore, after taking into account both theoretical requirements and experimental bounds, we find the allowed parameter space of the model. Our findings support the thought that the scalar boson found at the LHC with a mass 125 GeV could be dilaton dominance and it can be further addressed by the HL-LHC. Apart from that, we also give predictions of the four-point Higgs self-coupling for TSS and also HSS showing that both of them have weaker interactions than SM prediction. In the end, we compare the potential in our model with other models showing that it can give the same Higgs mass for $H_{125}$ based on different VEV and the distinct sharpness of TSS and HSS compared to the SM imply the thorough analysis of our model. \\

In this work, we consider a simple dilaton model and we can extend our results to other BSM models. In our analysis, we observe that the Yukawa couplings can constrain the model parameters strongly, i.e. most of parameter space of dilaton dominance can be excluded by $k_f$ measurements. \\
\indent
The results shown in Fig.~\ref{fig:fermions} are obviously model dependent, where we assume that dilaton field does not couple to fermions. Such results might be changed drastically if Yukawa couplings of dilaton (singlet) field and fermions are assumed, like the couplings of radion and fermions in Randall-Sundrum (RS) model \cite{Csaki:2007ns, Barger_2012} and the couplings of the singlet field with vector-like fermions in the Top quark see-saw mechanism \cite{He_2002,Abe:2012eu}. In these cases, larger parameter space of dilaton dominance might still be allowed by the LHC Higgs data. 
For example, the dilaton-dominant region for $s > 0.6$ in the TSS scenario might still be allowed if the couplings of dilaton and fermion fields are assumed. The couplings of the dilaton (singlet) field with fermions opens more potential theoretical interpretation of the observed 125 GeV Higgs boson $H_{125}$ as originating from the dilaton (singlet) scalar $\Phi$. At the same time, such couplings also offer theoretical potential to address more fundamental issues related to the neutrino masses and mixings, as shown in  \cite{Nishino:2004kb}. \\
\indent
The Weyl vector $\omega_\mu$, which becomes massive after scale-symmetry breaking, is naturally predicted to be a Higgs-portal dark-matter candidate \cite{PhysRevLett.61.2182,Huang:1989fj,Tang:2019uex}. In particular, it decouples from Higgs interactions at $\chi=0$ within the TSS framework.
 The gravitational waves might be produced by the phase transition accompanying with the conformal symmetry breaking \cite{ahriche2024gravitationalwavesphasetransitions}.
The model itself can be applied to interpret precision cosmology experiments, as demonstrated in \cite{Tang_2018, Ghilencea_2021}. Therefore, within this simple dilaton model, it is possible to shed light on the connection between particle physics and gravity.\\
\indent However, the existence of a set of parameters that can accommodate both cosmological and particle physics data deserves a careful study. 
At the current stage, there is an obvious tension in the SMW \cite{Ghilencea_2021, ghilencea_standard_2021}, where two sets of parameters for $\xi_h$ and $\xi$ in Eq.~\eqref{V_SMW} must be introduced in order to accommodate cosmological data (set B) and particle physics data (set A) as given in Table~\ref{tab:parameters_SMW}. The set A of parameters for EW spontaneous symmetry breaking used in Figure \ref{fig:potentials} is fixed at $\xi \sqrt{\xi_h}\sim 3.5\times 10^{-17}$ to produce the mass of $H_{125}$, which is significantly different from the cosmological set, where the set B of parameters requires that $\xi_h \sim 10^{-3}-10^{-2}$ and $\xi\sim 10^{-9}$ in order to accommodate inflation and CMB data. In term of $\xi \sqrt{\xi_h}$, the difference between these two sets of parameter can reach up to about 6–7 orders of magnitude, while in term of the scalar masses, the difference is 10 orders of magnitude. Such a tension deserves our further study.\\
\indent
The current work focuses on tree-level effects and collider phenomenological analysis, which remain meaningful at this stage. While in order to address the tension mentioned above, it might require to take into account inflation models and quantum corrections \cite{Donoghue:2017vvl, Tang_2018}. The solutions of this tension may shed light on the Higgs-related questions from the perspective of gravity and to investigate new physics using both collider and cosmology data.

Our model can give an explanation to the gravitational constant $G$ and the cosmological constant $\Lambda$, as in scalar-tensor theory \cite{faraoniCosmologyScalartensorGravity2004} after assuming $f\sim \Lambda_{\textrm{Pl}}$, because the model is constructed from a Weyl-gauge theory of gravity. However, the large hierarchy between the scales $f$ and $f_{EW}$ demands unnatural large parameters $\chi_{H,D}$ as seeing from Eqs. \eqref{Eff_Coup_Tri} and \eqref{Eff_Coup_hyp}. So, we would like to assume the scale $f$ is lower than $\Lambda_{\textrm{Pl}}$. Since we focus on the phenomenological analysis of the model in light of LHC measurements, the scale $f_{EW}$ is fixed while the scale symmetry breaking scale while the scale $f$ remains undetermined.

The last but not the least, it is necessary to address the issue of the radiative stability of the lighter scales in our model. As presented above, our model involves three fundamental physics scales, i.e. Planck energy scale $\Lambda_{\textrm{Pl}}$, conformal symmetry breaking scale $f$ and the electroweak breaking scale $f_{EW}$. It is natural to ask what mechanism can prevent the lighter scale, like $f_{EW}$, from being driven toward the heavier one, like $f$, by radiative corrections? Moreover, if the scale $f$ is distinct from the scale $\Lambda_{\textrm{Pl}}$, what prevents it from being pushed up to $\Lambda_{\textrm{Pl}}$ by quantum effects? In other words, when the relation $\Lambda_{\textrm{Pl}} \gg f \gg f_{EW}$ is assumed, this issue include two-fold aspects, the first aspect is why $f$ can not be driven up to $\Lambda_{\textrm{Pl}}$, and the second aspect is why $f_{EW}$ can not be driven up to $f$ by quantum corrections.

In order to address this issue, it is necessary to clarify scales in our model. At its face values, there are two sets of scales. The first set, ($f_H$, $f_D$), corresponds to the VEVs of the doublet scalar $\phi$ and dilaton $\Phi$ in the interaction basis. And the second set, ($f$, $f_{EW}$), corresponds to the physical scales associated with scale symmetry breaking and electroweak symmetry breaking. Due to the relation described in Eq.~\eqref{eq:dilaton_Higgs}, the first set of parameters determines the second set of the parameter after the symmetries breaking. 

It is noteworthy that our model offers a dynamic way to generate the mass term of the Higgs sector of the SM. For example, when the conformal symmetry is broken, saying $\Phi$ develops a vacuum $f_D$ for instance, the mass term of the SM Higgs given in Eq.~\eqref{V_SM} can be generated, which reads 
 \begin{equation}
 \mu^2 = - \frac{\alpha}{2} f_D^2
 \end{equation}
 The sign of $\alpha$ must be negative in order to have a nonvanishing $f_H$. Since $f$ is the physical scale for the gauged scale symmetry breaking and, as shown in Eqs. \eqref{Eff_Coup_Tri} and \eqref{Eff_Coup_hyp}, $f_{EW}$ is proportional to $f$ by the following relation
\begin{equation}
f_{EW}  \approx   f \sqrt{ \frac{- \alpha + \frac{\rho \chi_H^2 }{3 \chi_D^2} }{2 \chi_D^2 (\lambda + \frac{\rho \chi_H^4}{6 \chi_D^4} - \frac{\alpha \chi_H^2}{\chi_D^2})} }     
\end{equation}
if $\frac{\chi_H}{f} U_t$ is a small number in the TSS, or by the relation
\begin{equation}
f_{EW}  \approx  f \sqrt{ \frac{ - \alpha - \frac{\rho}{3} \frac{\chi_H^2}{\chi_D^2} }{2 \chi_D^2 (\lambda + \alpha \frac{\chi_H^2}{\chi_D^2} + \frac{\rho}{6} \frac{\chi_H^4}{\chi_D^4}) } }     
\end{equation}
if $\frac{\chi_H}{f} U_h$ is a small number in the HSS. It is observed that these two physical scales, $f$ and $f_{EW}$, are connected with each other in the model.

On the first aspect of the issue, if $\Lambda_{\textrm{Pl}} \gg f$, whether the radiative corrections can drive $f$ up to $\Lambda_{\textrm{Pl}}$? The answer is no. For example, in composite Higgs models, the scale $f$ arises from gluon condensates \cite{migdal_dilaton_1982} or fermion condensates \cite{Weinberg:1975gm,zwicky_qcd_2023, appelquist_dilaton_2022}. If the scale $f$ is much smaller than $\Lambda_{\textrm{Pl}}$, above the scale $f$, the conformal symmetry can work and guarantee $f$ is only a low energy scale and can keep $f$ being smaller than $\Lambda_{\textrm{Pl}}$. Supposing the dilaton field is a low energy degree of freedom from gluon condensates, like in the setup of reference \cite{migdal_dilaton_1982}. Above the scale $f$, the low energy effective theory of dilaton field turns into a pure Yang-Mills theory which respects the conformal symmetry exactly, where the scale $f$ is a low energy scale and might be directly related with the dimensional transmutation of strong Yang-Mills dynamics.

The second aspect of the issue is more subtle. If two physical scales $f$ and $f_{EW}$ are close to each other, saying $f \sim f_{EW}$, there is no need to worry about the radiative stability of the model. 

While for the case $f \gg f_{EW}$ (saying $f = 10^{3-15} f_{EW}$), we have to face with the issue of radiative stability for the lighter scale or our model will suffer the little/big hierarchy problem. Since below $f$ there is no remaining conformal symmetry to protect the mass term $ \mu^2$ from getting a large quantum correction, it is expected that the quantum correction of $\mu^2$ should be $\mu^2 + \delta \mu^2 \propto f_{EW}^2 + C\,f^2$ (where $C$ is related to model parameters), just like the quadratic divergence of Higgs mass term in the SM.

The issue of radiative stability of $f_{EW}$ can be investigated in the holographic RS model. In RS model \cite{Randall:1999ee}, the effective scales $f \sim \Lambda_{\textrm{Pl}}$ and $f_{EW}$, arise from UV and IR branes, respectively. A bulk scalar, called as the radion, is introduced to stabilize the warped geometry of the extra dimension \cite{Goldberger:1999uk}. The hierarchy problem is translated into the warp factor between the UV and IR branes and is controlled by the radion.  Within the holographic framework, the duality between the radion and the dilaton introduces a scale symmetry that can prevent the scale $f_{EW}$ from being pushed to $\Lambda_{\textrm{Pl}}$ by quantum corrections \cite{Frampton:1999yb} as long as the scaling dimension $\Delta_\mathcal{O}$ approaches the canonical dimension $d_{\mathcal{O}}$ \cite{Rattazzi:2000hs}, which gives rise to the Coleman-Weinberg correction \cite{Goldberger:2007zk}. Other plausible mechanisms in literature \cite{Csaki:2015hcd, Csaki:2018muy,Csaki:2021abz}, such as an approximate shift symmetry for the light scalar \cite{Graham:2015cka}, compositeness \cite{Ferretti:2021jai}, or near-conformal dynamics operative below $f$ \cite{Cata:2018wzl,Crewther:2020tgd}, could protect the lighter scale without qualitatively altering the phenomenological studied in this work.

Therefore, we can conclude that an exact or approximate gauged scale symmetry above the scale $f$ can keeps the scale $f$ to be stable if $\Lambda_{\textrm{Pl}} \gg f$, while additional dynamics might be needed if $f \gg f_{EW}$ in order to keep the lower scale $f_{EW}$ stable. In our phenomenological analysis, we have fixed $f_{EW}$ and have left $f$ as a free parameter. Thus our results is independent of the actual stabilizing mechanism for the lower scale $f_{EW}$.

\begin{acknowledgments} 
We thank Sichun Sun, Yong Tang and Tianrui Che for useful discussions. This work is supported by the Natural Science Foundation of China under the Grants No. 11875260 and No. 12275143.
\end{acknowledgments} 

%

\twocolumngrid

\bibliography{references}

\begin{thebibliography}{83}%
\makeatletter
\providecommand \@ifxundefined [1]{%
 \@ifx{#1\undefined}
}%
\providecommand \@ifnum [1]{%
 \ifnum #1\expandafter \@firstoftwo
 \else \expandafter \@secondoftwo
 \fi
}%
\providecommand \@ifx [1]{%
 \ifx #1\expandafter \@firstoftwo
 \else \expandafter \@secondoftwo
 \fi
}%
\providecommand \natexlab [1]{#1}%
\providecommand \enquote  [1]{``#1''}%
\providecommand \bibnamefont  [1]{#1}%
\providecommand \bibfnamefont [1]{#1}%
\providecommand \citenamefont [1]{#1}%
\providecommand \href@noop [0]{\@secondoftwo}%
\providecommand \href [0]{\begingroup \@sanitize@url \@href}%
\providecommand \@href[1]{\@@startlink{#1}\@@href}%
\providecommand \@@href[1]{\endgroup#1\@@endlink}%
\providecommand \@sanitize@url [0]{\catcode `\\12\catcode `\$12\catcode `\&12\catcode `\#12\catcode `\^12\catcode `\_12\catcode `\%12\relax}%
\providecommand \@@startlink[1]{}%
\providecommand \@@endlink[0]{}%
\providecommand \url  [0]{\begingroup\@sanitize@url \@url }%
\providecommand \@url [1]{\endgroup\@href {#1}{\urlprefix }}%
\providecommand \urlprefix  [0]{URL }%
\providecommand \Eprint [0]{\href }%
\providecommand \doibase [0]{http://dx.doi.org/}%
\providecommand \selectlanguage [0]{\@gobble}%
\providecommand \bibinfo  [0]{\@secondoftwo}%
\providecommand \bibfield  [0]{\@secondoftwo}%
\providecommand \translation [1]{[#1]}%
\providecommand \BibitemOpen [0]{}%
\providecommand \bibitemStop [0]{}%
\providecommand \bibitemNoStop [0]{.\EOS\space}%
\providecommand \EOS [0]{\spacefactor3000\relax}%
\providecommand \BibitemShut  [1]{\csname bibitem#1\endcsname}%
\let\auto@bib@innerbib\@empty
\bibitem [{\citenamefont {Aad}\ \emph {et~al.}(2012)\citenamefont {Aad} \emph {et~al.}}]{ATLAS:2012yve}%
  \BibitemOpen
  \bibfield  {author} {\bibinfo {author} {\bibfnamefont {Georges}\ \bibnamefont {Aad}} \emph {et~al.} (\bibinfo {collaboration} {ATLAS}),\ }\bibfield  {title} {\enquote {\bibinfo {title} {{Observation of a new particle in the search for the Standard Model Higgs boson with the ATLAS detector at the LHC}},}\ }\href {\doibase 10.1016/j.physletb.2012.08.020} {\bibfield  {journal} {\bibinfo  {journal} {Phys. Lett. B}\ }\textbf {\bibinfo {volume} {716}},\ \bibinfo {pages} {1--29} (\bibinfo {year} {2012})},\ \Eprint {http://arxiv.org/abs/1207.7214} {arXiv:1207.7214 [hep-ex]} \BibitemShut {NoStop}%
\bibitem [{\citenamefont {Chatrchyan}\ \emph {et~al.}(2012)\citenamefont {Chatrchyan} \emph {et~al.}}]{CMS:2012qbp}%
  \BibitemOpen
  \bibfield  {author} {\bibinfo {author} {\bibfnamefont {Serguei}\ \bibnamefont {Chatrchyan}} \emph {et~al.} (\bibinfo {collaboration} {CMS}),\ }\bibfield  {title} {\enquote {\bibinfo {title} {{Observation of a New Boson at a Mass of 125 GeV with the CMS Experiment at the LHC}},}\ }\href {\doibase 10.1016/j.physletb.2012.08.021} {\bibfield  {journal} {\bibinfo  {journal} {Phys. Lett. B}\ }\textbf {\bibinfo {volume} {716}},\ \bibinfo {pages} {30--61} (\bibinfo {year} {2012})},\ \Eprint {http://arxiv.org/abs/1207.7235} {arXiv:1207.7235 [hep-ex]} \BibitemShut {NoStop}%
\bibitem [{\citenamefont {Abbott}\ \emph {et~al.}(2016)\citenamefont {Abbott} \emph {et~al.}}]{LIGOScientific:2016aoc}%
  \BibitemOpen
  \bibfield  {author} {\bibinfo {author} {\bibfnamefont {B.~P.}\ \bibnamefont {Abbott}} \emph {et~al.} (\bibinfo {collaboration} {LIGO Scientific, Virgo}),\ }\bibfield  {title} {\enquote {\bibinfo {title} {{Observation of Gravitational Waves from a Binary Black Hole Merger}},}\ }\href {\doibase 10.1103/PhysRevLett.116.061102} {\bibfield  {journal} {\bibinfo  {journal} {Phys. Rev. Lett.}\ }\textbf {\bibinfo {volume} {116}},\ \bibinfo {pages} {061102} (\bibinfo {year} {2016})},\ \Eprint {http://arxiv.org/abs/1602.03837} {arXiv:1602.03837 [gr-qc]} \BibitemShut {NoStop}%
\bibitem [{\citenamefont {Stelle}(1977)}]{Stelle:1976gc}%
  \BibitemOpen
  \bibfield  {author} {\bibinfo {author} {\bibfnamefont {K.~S.}\ \bibnamefont {Stelle}},\ }\bibfield  {title} {\enquote {\bibinfo {title} {{Renormalization of Higher Derivative Quantum Gravity}},}\ }\href {\doibase 10.1103/PhysRevD.16.953} {\bibfield  {journal} {\bibinfo  {journal} {Phys. Rev. D}\ }\textbf {\bibinfo {volume} {16}},\ \bibinfo {pages} {953--969} (\bibinfo {year} {1977})}\BibitemShut {NoStop}%
\bibitem [{\citenamefont {Horava}(2009)}]{Horava:2009uw}%
  \BibitemOpen
  \bibfield  {author} {\bibinfo {author} {\bibfnamefont {Petr}\ \bibnamefont {Horava}},\ }\bibfield  {title} {\enquote {\bibinfo {title} {{Quantum Gravity at a Lifshitz Point}},}\ }\href {\doibase 10.1103/PhysRevD.79.084008} {\bibfield  {journal} {\bibinfo  {journal} {Phys. Rev. D}\ }\textbf {\bibinfo {volume} {79}},\ \bibinfo {pages} {084008} (\bibinfo {year} {2009})},\ \Eprint {http://arxiv.org/abs/0901.3775} {arXiv:0901.3775 [hep-th]} \BibitemShut {NoStop}%
\bibitem [{\citenamefont {Modesto}(2012)}]{Modesto:2011kw}%
  \BibitemOpen
  \bibfield  {author} {\bibinfo {author} {\bibfnamefont {Leonardo}\ \bibnamefont {Modesto}},\ }\bibfield  {title} {\enquote {\bibinfo {title} {{Super-renormalizable Quantum Gravity}},}\ }\href {\doibase 10.1103/PhysRevD.86.044005} {\bibfield  {journal} {\bibinfo  {journal} {Phys. Rev. D}\ }\textbf {\bibinfo {volume} {86}},\ \bibinfo {pages} {044005} (\bibinfo {year} {2012})},\ \Eprint {http://arxiv.org/abs/1107.2403} {arXiv:1107.2403 [hep-th]} \BibitemShut {NoStop}%
\bibitem [{\citenamefont {Hayashi}\ \emph {et~al.}(1978)\citenamefont {Hayashi}, \citenamefont {Kasuya},\ and\ \citenamefont {Shirafuji}}]{hayashi_elementary_1977}%
  \BibitemOpen
  \bibfield  {author} {\bibinfo {author} {\bibfnamefont {Kenji}\ \bibnamefont {Hayashi}}, \bibinfo {author} {\bibfnamefont {Masahiro}\ \bibnamefont {Kasuya}}, \ and\ \bibinfo {author} {\bibfnamefont {Takeshi}\ \bibnamefont {Shirafuji}},\ }\bibfield  {title} {\enquote {\bibinfo {title} {Elementary particles and weyl's gauge field},}\ }\href {\doibase 10.1143/PTP.59.681} {\bibfield  {journal} {\bibinfo  {journal} {Progress of Theoretical Physics}\ }\textbf {\bibinfo {volume} {59}},\ \bibinfo {pages} {681--681} (\bibinfo {year} {1978})},\ \Eprint {http://arxiv.org/abs/https://academic.oup.com/ptp/article-pdf/59/2/681/5308006/59-2-681.pdf} {https://academic.oup.com/ptp/article-pdf/59/2/681/5308006/59-2-681.pdf} \BibitemShut {NoStop}%
\bibitem [{\citenamefont {{Hayashi}}\ and\ \citenamefont {{Kugo}}(1979)}]{hayashi_remarks_1979}%
  \BibitemOpen
  \bibfield  {author} {\bibinfo {author} {\bibfnamefont {K.}~\bibnamefont {{Hayashi}}}\ and\ \bibinfo {author} {\bibfnamefont {T.}~\bibnamefont {{Kugo}}},\ }\bibfield  {title} {\enquote {\bibinfo {title} {{Remarks on Weyl's Gauge Field}},}\ }\href {\doibase 10.1143/PTP.61.334} {\bibfield  {journal} {\bibinfo  {journal} {Progress of Theoretical Physics}\ }\textbf {\bibinfo {volume} {61}},\ \bibinfo {pages} {334--346} (\bibinfo {year} {1979})}\BibitemShut {NoStop}%
\bibitem [{\citenamefont {Ghilencea}\ and\ \citenamefont {Lee}(2019)}]{ghilencea_Weyl_2019}%
  \BibitemOpen
  \bibfield  {author} {\bibinfo {author} {\bibfnamefont {D.~M.}\ \bibnamefont {Ghilencea}}\ and\ \bibinfo {author} {\bibfnamefont {Hyun~Min}\ \bibnamefont {Lee}},\ }\bibfield  {title} {\enquote {\bibinfo {title} {{Weyl gauge symmetry and its spontaneous breaking in the standard model and inflation}},}\ }\href {\doibase 10.1103/PhysRevD.99.115007} {\bibfield  {journal} {\bibinfo  {journal} {Phys. Rev. D}\ }\textbf {\bibinfo {volume} {99}},\ \bibinfo {pages} {115007} (\bibinfo {year} {2019})},\ \Eprint {http://arxiv.org/abs/1809.09174} {arXiv:1809.09174 [hep-th]} \BibitemShut {NoStop}%
\bibitem [{\citenamefont {Ghilencea}(2019)}]{ghilencea_spontaneous_2019}%
  \BibitemOpen
  \bibfield  {author} {\bibinfo {author} {\bibfnamefont {D.~M.}\ \bibnamefont {Ghilencea}},\ }\bibfield  {title} {\enquote {\bibinfo {title} {Spontaneous breaking of weyl quadratic gravity to einstein action and higgs potential},}\ }\href {\doibase 10.1007/jhep03(2019)049} {\bibfield  {journal} {\bibinfo  {journal} {Journal of High Energy Physics}\ }\textbf {\bibinfo {volume} {2019}} (\bibinfo {year} {2019}),\ 10.1007/jhep03(2019)049}\BibitemShut {NoStop}%
\bibitem [{\citenamefont {Ghilencea}(2020)}]{ghilencea_stueckelberg_2020}%
  \BibitemOpen
  \bibfield  {author} {\bibinfo {author} {\bibfnamefont {D.~M.}\ \bibnamefont {Ghilencea}},\ }\bibfield  {title} {\enquote {\bibinfo {title} {Stueckelberg breaking of weyl conformal geometry and applications to gravity},}\ }\href {\doibase 10.1103/PhysRevD.101.045010} {\bibfield  {journal} {\bibinfo  {journal} {Phys. Rev. D}\ }\textbf {\bibinfo {volume} {101}},\ \bibinfo {pages} {045010} (\bibinfo {year} {2020})}\BibitemShut {NoStop}%
\bibitem [{\citenamefont {Ghilencea}(2022)}]{ghilencea_standard_2021}%
  \BibitemOpen
  \bibfield  {author} {\bibinfo {author} {\bibfnamefont {D.~M.}\ \bibnamefont {Ghilencea}},\ }\bibfield  {title} {\enquote {\bibinfo {title} {{Standard Model in Weyl conformal geometry}},}\ }\href {\doibase 10.1140/epjc/s10052-021-09887-y} {\bibfield  {journal} {\bibinfo  {journal} {Eur. Phys. J. C}\ }\textbf {\bibinfo {volume} {82}},\ \bibinfo {pages} {23} (\bibinfo {year} {2022})},\ \Eprint {http://arxiv.org/abs/2104.15118} {arXiv:2104.15118 [hep-ph]} \BibitemShut {NoStop}%
\bibitem [{\citenamefont {Ghilencea}(2023)}]{ghilencea_non-metric_2023}%
  \BibitemOpen
  \bibfield  {author} {\bibinfo {author} {\bibfnamefont {D.~M.}\ \bibnamefont {Ghilencea}},\ }\bibfield  {title} {\enquote {\bibinfo {title} {Non-metric geometry as the origin of mass in gauge theories of scale invariance},}\ }\href {\doibase 10.1140/epjc/s10052-023-11237-z} {\bibfield  {journal} {\bibinfo  {journal} {The European Physical Journal C}\ }\textbf {\bibinfo {volume} {83}} (\bibinfo {year} {2023}),\ 10.1140/epjc/s10052-023-11237-z}\BibitemShut {NoStop}%
\bibitem [{\citenamefont {Ghilencea}\ and\ \citenamefont {Hill}(2024)}]{ghilencea_standard_2024}%
  \BibitemOpen
  \bibfield  {author} {\bibinfo {author} {\bibfnamefont {D.~M.}\ \bibnamefont {Ghilencea}}\ and\ \bibinfo {author} {\bibfnamefont {C.~T.}\ \bibnamefont {Hill}},\ }\bibfield  {title} {\enquote {\bibinfo {title} {{Standard Model in conformal geometry: Local vs gauged scale invariance}},}\ }\href {\doibase 10.1016/j.aop.2023.169562} {\bibfield  {journal} {\bibinfo  {journal} {Annals Phys.}\ }\textbf {\bibinfo {volume} {460}},\ \bibinfo {pages} {169562} (\bibinfo {year} {2024})},\ \Eprint {http://arxiv.org/abs/2303.02515} {arXiv:2303.02515 [hep-th]} \BibitemShut {NoStop}%
\bibitem [{\citenamefont {Weyl}(1918)}]{Weyl:1918ib}%
  \BibitemOpen
  \bibfield  {author} {\bibinfo {author} {\bibfnamefont {H.}~\bibnamefont {Weyl}},\ }\bibfield  {title} {\enquote {\bibinfo {title} {{Gravitation and electricity}},}\ }\href@noop {} {\bibfield  {journal} {\bibinfo  {journal} {Sitzungsber. Preuss. Akad. Wiss. Berlin (Math. Phys. )}\ }\textbf {\bibinfo {volume} {1918}},\ \bibinfo {pages} {465} (\bibinfo {year} {1918})}\BibitemShut {NoStop}%
\bibitem [{\citenamefont {Dirac}(1973)}]{Dirac:1973gk}%
  \BibitemOpen
  \bibfield  {author} {\bibinfo {author} {\bibfnamefont {Paul A.~M.}\ \bibnamefont {Dirac}},\ }\bibfield  {title} {\enquote {\bibinfo {title} {{Long range forces and broken symmetries}},}\ }\href {\doibase 10.1098/rspa.1973.0070} {\bibfield  {journal} {\bibinfo  {journal} {Proc. Roy. Soc. Lond. A}\ }\textbf {\bibinfo {volume} {333}},\ \bibinfo {pages} {403--418} (\bibinfo {year} {1973})}\BibitemShut {NoStop}%
\bibitem [{\citenamefont {O'Raifeartaigh}(1997)}]{ORaifeartaigh:1997dvq}%
  \BibitemOpen
  \bibfield  {author} {\bibinfo {author} {\bibfnamefont {L.}~\bibnamefont {O'Raifeartaigh}},\ }\href@noop {} {\emph {\bibinfo {title} {{The dawning of gauge theory}}}}\ (\bibinfo  {publisher} {Princeton Univ. Press},\ \bibinfo {address} {Princeton, NJ, USA},\ \bibinfo {year} {1997})\BibitemShut {NoStop}%
\bibitem [{\citenamefont {Fujii}(1982)}]{Fujii:1982ms}%
  \BibitemOpen
  \bibfield  {author} {\bibinfo {author} {\bibfnamefont {Yasunori}\ \bibnamefont {Fujii}},\ }\bibfield  {title} {\enquote {\bibinfo {title} {{Origin of the Gravitational Constant and Particle Masses in Scale Invariant Scalar - Tensor Theory}},}\ }\href {\doibase 10.1103/PhysRevD.26.2580} {\bibfield  {journal} {\bibinfo  {journal} {Phys. Rev. D}\ }\textbf {\bibinfo {volume} {26}},\ \bibinfo {pages} {2580} (\bibinfo {year} {1982})}\BibitemShut {NoStop}%
\bibitem [{\citenamefont {Drechsler}\ and\ \citenamefont {Tann}(1999)}]{Drechsler:1998gy}%
  \BibitemOpen
  \bibfield  {author} {\bibinfo {author} {\bibfnamefont {W.}~\bibnamefont {Drechsler}}\ and\ \bibinfo {author} {\bibfnamefont {H.}~\bibnamefont {Tann}},\ }\bibfield  {title} {\enquote {\bibinfo {title} {{Broken Weyl invariance and the origin of mass}},}\ }\href {\doibase 10.1023/A:1012851715278} {\bibfield  {journal} {\bibinfo  {journal} {Found. Phys.}\ }\textbf {\bibinfo {volume} {29}},\ \bibinfo {pages} {1023--1064} (\bibinfo {year} {1999})},\ \Eprint {http://arxiv.org/abs/gr-qc/9802044} {arXiv:gr-qc/9802044} \BibitemShut {NoStop}%
\bibitem [{\citenamefont {Chamseddine}\ \emph {et~al.}(2007)\citenamefont {Chamseddine}, \citenamefont {Connes},\ and\ \citenamefont {Marcolli}}]{Chamseddine:2006ep}%
  \BibitemOpen
  \bibfield  {author} {\bibinfo {author} {\bibfnamefont {Ali~H.}\ \bibnamefont {Chamseddine}}, \bibinfo {author} {\bibfnamefont {Alain}\ \bibnamefont {Connes}}, \ and\ \bibinfo {author} {\bibfnamefont {Matilde}\ \bibnamefont {Marcolli}},\ }\bibfield  {title} {\enquote {\bibinfo {title} {{Gravity and the standard model with neutrino mixing}},}\ }\href {\doibase 10.4310/ATMP.2007.v11.n6.a3} {\bibfield  {journal} {\bibinfo  {journal} {Adv. Theor. Math. Phys.}\ }\textbf {\bibinfo {volume} {11}},\ \bibinfo {pages} {991--1089} (\bibinfo {year} {2007})},\ \Eprint {http://arxiv.org/abs/hep-th/0610241} {arXiv:hep-th/0610241} \BibitemShut {NoStop}%
\bibitem [{\citenamefont {Kaluza}(1921)}]{Kaluza:1921tu}%
  \BibitemOpen
  \bibfield  {author} {\bibinfo {author} {\bibfnamefont {Th.}\ \bibnamefont {Kaluza}},\ }\bibfield  {title} {\enquote {\bibinfo {title} {{Zum Unit\"atsproblem der Physik}},}\ }\href {\doibase 10.1142/S0218271818700017} {\bibfield  {journal} {\bibinfo  {journal} {Sitzungsber. Preuss. Akad. Wiss. Berlin (Math. Phys. )}\ }\textbf {\bibinfo {volume} {1921}},\ \bibinfo {pages} {966--972} (\bibinfo {year} {1921})},\ \Eprint {http://arxiv.org/abs/1803.08616} {arXiv:1803.08616 [physics.hist-ph]} \BibitemShut {NoStop}%
\bibitem [{\citenamefont {Klein}(1926)}]{Klein:1926tv}%
  \BibitemOpen
  \bibfield  {author} {\bibinfo {author} {\bibfnamefont {Oskar}\ \bibnamefont {Klein}},\ }\bibfield  {title} {\enquote {\bibinfo {title} {{Quantum Theory and Five-Dimensional Theory of Relativity. (In German and English)}},}\ }\href {\doibase 10.1007/BF01397481} {\bibfield  {journal} {\bibinfo  {journal} {Z. Phys.}\ }\textbf {\bibinfo {volume} {37}},\ \bibinfo {pages} {895--906} (\bibinfo {year} {1926})}\BibitemShut {NoStop}%
\bibitem [{\citenamefont {Becker}\ \emph {et~al.}(2006)\citenamefont {Becker}, \citenamefont {Becker},\ and\ \citenamefont {Schwarz}}]{Becker:2006dvp}%
  \BibitemOpen
  \bibfield  {author} {\bibinfo {author} {\bibfnamefont {K.}~\bibnamefont {Becker}}, \bibinfo {author} {\bibfnamefont {M.}~\bibnamefont {Becker}}, \ and\ \bibinfo {author} {\bibfnamefont {J.~H.}\ \bibnamefont {Schwarz}},\ }\href {\doibase 10.1017/CBO9780511816086} {\emph {\bibinfo {title} {{String theory and M-theory: A modern introduction}}}}\ (\bibinfo  {publisher} {Cambridge University Press},\ \bibinfo {year} {2006})\BibitemShut {NoStop}%
\bibitem [{\citenamefont {Barger}\ \emph {et~al.}(2012)\citenamefont {Barger}, \citenamefont {Ishida},\ and\ \citenamefont {Keung}}]{Barger_2012}%
  \BibitemOpen
  \bibfield  {author} {\bibinfo {author} {\bibfnamefont {Vernon}\ \bibnamefont {Barger}}, \bibinfo {author} {\bibfnamefont {Muneyuki}\ \bibnamefont {Ishida}}, \ and\ \bibinfo {author} {\bibfnamefont {Wai-Yee}\ \bibnamefont {Keung}},\ }\bibfield  {title} {\enquote {\bibinfo {title} {Differentiating the higgs boson from the dilaton and radion at hadron colliders},}\ }\href {\doibase 10.1103/physrevlett.108.101802} {\bibfield  {journal} {\bibinfo  {journal} {Physical Review Letters}\ }\textbf {\bibinfo {volume} {108}} (\bibinfo {year} {2012}),\ 10.1103/physrevlett.108.101802}\BibitemShut {NoStop}%
\bibitem [{\citenamefont {Matsuzaki}\ and\ \citenamefont {Yamawaki}(2012)}]{matsuzaki_discovering_2012}%
  \BibitemOpen
  \bibfield  {author} {\bibinfo {author} {\bibfnamefont {Shinya}\ \bibnamefont {Matsuzaki}}\ and\ \bibinfo {author} {\bibfnamefont {Koichi}\ \bibnamefont {Yamawaki}},\ }\bibfield  {title} {\enquote {\bibinfo {title} {Discovering the 125 gev techni-dilaton at the lhc},}\ }\href {\doibase 10.1103/physrevd.86.035025} {\bibfield  {journal} {\bibinfo  {journal} {Physical Review D}\ }\textbf {\bibinfo {volume} {86}} (\bibinfo {year} {2012}),\ 10.1103/physrevd.86.035025}\BibitemShut {NoStop}%
\bibitem [{\citenamefont {Bellazzini}\ \emph {et~al.}(2013)\citenamefont {Bellazzini}, \citenamefont {Csáki}, \citenamefont {Hubisz}, \citenamefont {Serra},\ and\ \citenamefont {Terning}}]{Bellazzini_2013}%
  \BibitemOpen
  \bibfield  {author} {\bibinfo {author} {\bibfnamefont {Brando}\ \bibnamefont {Bellazzini}}, \bibinfo {author} {\bibfnamefont {Csaba}\ \bibnamefont {Csáki}}, \bibinfo {author} {\bibfnamefont {Jay}\ \bibnamefont {Hubisz}}, \bibinfo {author} {\bibfnamefont {Javi}\ \bibnamefont {Serra}}, \ and\ \bibinfo {author} {\bibfnamefont {John}\ \bibnamefont {Terning}},\ }\bibfield  {title} {\enquote {\bibinfo {title} {A higgs-like dilaton},}\ }\href {\doibase 10.1140/epjc/s10052-013-2333-x} {\bibfield  {journal} {\bibinfo  {journal} {The European Physical Journal C}\ }\textbf {\bibinfo {volume} {73}} (\bibinfo {year} {2013}),\ 10.1140/epjc/s10052-013-2333-x}\BibitemShut {NoStop}%
\bibitem [{\citenamefont {Abe}\ \emph {et~al.}(2012)\citenamefont {Abe}, \citenamefont {Kitano}, \citenamefont {Konishi}, \citenamefont {Oda}, \citenamefont {Sato},\ and\ \citenamefont {Sugiyama}}]{Abe:2012eu}%
  \BibitemOpen
  \bibfield  {author} {\bibinfo {author} {\bibfnamefont {Tomohiro}\ \bibnamefont {Abe}}, \bibinfo {author} {\bibfnamefont {Ryuichiro}\ \bibnamefont {Kitano}}, \bibinfo {author} {\bibfnamefont {Yasufumi}\ \bibnamefont {Konishi}}, \bibinfo {author} {\bibfnamefont {Kin-ya}\ \bibnamefont {Oda}}, \bibinfo {author} {\bibfnamefont {Joe}\ \bibnamefont {Sato}}, \ and\ \bibinfo {author} {\bibfnamefont {Shohei}\ \bibnamefont {Sugiyama}},\ }\bibfield  {title} {\enquote {\bibinfo {title} {{Minimal Dilaton Model}},}\ }\href {\doibase 10.1103/PhysRevD.86.115016} {\bibfield  {journal} {\bibinfo  {journal} {Phys. Rev. D}\ }\textbf {\bibinfo {volume} {86}},\ \bibinfo {pages} {115016} (\bibinfo {year} {2012})},\ \Eprint {http://arxiv.org/abs/1209.4544} {arXiv:1209.4544 [hep-ph]} \BibitemShut {NoStop}%
\bibitem [{\citenamefont {Flato}\ and\ \citenamefont {Raczka}(1988)}]{Flato:1987bb}%
  \BibitemOpen
  \bibfield  {author} {\bibinfo {author} {\bibfnamefont {Moshe}\ \bibnamefont {Flato}}\ and\ \bibinfo {author} {\bibfnamefont {Ryszard}\ \bibnamefont {Raczka}},\ }\bibfield  {title} {\enquote {\bibinfo {title} {{A Possible Gravitational Origin of Higgs Field in the Standard Model}},}\ }\href {\doibase 10.1016/0370-2693(88)91213-0} {\bibfield  {journal} {\bibinfo  {journal} {Phys. Lett. B}\ }\textbf {\bibinfo {volume} {208}},\ \bibinfo {pages} {110--114} (\bibinfo {year} {1988})}\BibitemShut {NoStop}%
\bibitem [{\citenamefont {van~der Bij}(1994)}]{vanderBij:1993hx}%
  \BibitemOpen
  \bibfield  {author} {\bibinfo {author} {\bibfnamefont {J.~J.}\ \bibnamefont {van~der Bij}},\ }\bibfield  {title} {\enquote {\bibinfo {title} {{Can gravity make the Higgs particle decouple?}}}\ }\href@noop {} {\bibfield  {journal} {\bibinfo  {journal} {Acta Phys. Polon. B}\ }\textbf {\bibinfo {volume} {25}},\ \bibinfo {pages} {827--832} (\bibinfo {year} {1994})}\BibitemShut {NoStop}%
\bibitem [{\citenamefont {Cheng}(1988)}]{PhysRevLett.61.2182}%
  \BibitemOpen
  \bibfield  {author} {\bibinfo {author} {\bibfnamefont {Hung}\ \bibnamefont {Cheng}},\ }\bibfield  {title} {\enquote {\bibinfo {title} {Possible existence of weyl's vector meson},}\ }\href {\doibase 10.1103/PhysRevLett.61.2182} {\bibfield  {journal} {\bibinfo  {journal} {Phys. Rev. Lett.}\ }\textbf {\bibinfo {volume} {61}},\ \bibinfo {pages} {2182--2184} (\bibinfo {year} {1988})}\BibitemShut {NoStop}%
\bibitem [{\citenamefont {Nishino}\ and\ \citenamefont {Rajpoot}(2004)}]{Nishino:2004kb}%
  \BibitemOpen
  \bibfield  {author} {\bibinfo {author} {\bibfnamefont {Hitoshi}\ \bibnamefont {Nishino}}\ and\ \bibinfo {author} {\bibfnamefont {Subhash}\ \bibnamefont {Rajpoot}},\ }\bibfield  {title} {\enquote {\bibinfo {title} {{Broken scale invariance in the standard model}},}\ }\href@noop {} {\  (\bibinfo {year} {2004})},\ \Eprint {http://arxiv.org/abs/hep-th/0403039} {arXiv:hep-th/0403039} \BibitemShut {NoStop}%
\bibitem [{\citenamefont {Nishino}\ and\ \citenamefont {Rajpoot}(2009)}]{Nishino:2009in}%
  \BibitemOpen
  \bibfield  {author} {\bibinfo {author} {\bibfnamefont {Hitoshi}\ \bibnamefont {Nishino}}\ and\ \bibinfo {author} {\bibfnamefont {Subhash}\ \bibnamefont {Rajpoot}},\ }\bibfield  {title} {\enquote {\bibinfo {title} {{Implication of Compensator Field and Local Scale Invariance in the Standard Model}},}\ }\href {\doibase 10.1103/PhysRevD.79.125025} {\bibfield  {journal} {\bibinfo  {journal} {Phys. Rev. D}\ }\textbf {\bibinfo {volume} {79}},\ \bibinfo {pages} {125025} (\bibinfo {year} {2009})},\ \Eprint {http://arxiv.org/abs/0906.4778} {arXiv:0906.4778 [hep-th]} \BibitemShut {NoStop}%
\bibitem [{\citenamefont {de~Cesare}\ \emph {et~al.}(2017)\citenamefont {de~Cesare}, \citenamefont {Moffat},\ and\ \citenamefont {Sakellariadou}}]{deCesare:2016mml}%
  \BibitemOpen
  \bibfield  {author} {\bibinfo {author} {\bibfnamefont {Marco}\ \bibnamefont {de~Cesare}}, \bibinfo {author} {\bibfnamefont {John~W.}\ \bibnamefont {Moffat}}, \ and\ \bibinfo {author} {\bibfnamefont {Mairi}\ \bibnamefont {Sakellariadou}},\ }\bibfield  {title} {\enquote {\bibinfo {title} {{Local conformal symmetry in non-Riemannian geometry and the origin of physical scales}},}\ }\href {\doibase 10.1140/epjc/s10052-017-5183-0} {\bibfield  {journal} {\bibinfo  {journal} {Eur. Phys. J. C}\ }\textbf {\bibinfo {volume} {77}},\ \bibinfo {pages} {605} (\bibinfo {year} {2017})},\ \Eprint {http://arxiv.org/abs/1612.08066} {arXiv:1612.08066 [hep-th]} \BibitemShut {NoStop}%
\bibitem [{\citenamefont {Hehl}\ \emph {et~al.}(1995)\citenamefont {Hehl}, \citenamefont {McCrea}, \citenamefont {Mielke},\ and\ \citenamefont {Ne’eman}}]{hehl_metric_1995}%
  \BibitemOpen
  \bibfield  {author} {\bibinfo {author} {\bibfnamefont {Friedrich~W.}\ \bibnamefont {Hehl}}, \bibinfo {author} {\bibfnamefont {J.Dermott}\ \bibnamefont {McCrea}}, \bibinfo {author} {\bibfnamefont {Eckehard~W.}\ \bibnamefont {Mielke}}, \ and\ \bibinfo {author} {\bibfnamefont {Yuval}\ \bibnamefont {Ne’eman}},\ }\bibfield  {title} {\enquote {\bibinfo {title} {Metric-affine gauge theory of gravity: field equations, noether identities, world spinors, and breaking of dilation invariance},}\ }\href {\doibase 10.1016/0370-1573(94)00111-f} {\bibfield  {journal} {\bibinfo  {journal} {Physics Reports}\ }\textbf {\bibinfo {volume} {258}},\ \bibinfo {pages} {1–171} (\bibinfo {year} {1995})}\BibitemShut {NoStop}%
\bibitem [{\citenamefont {Trautman}(1979)}]{trautman_geometry_1979}%
  \BibitemOpen
  \bibfield  {author} {\bibinfo {author} {\bibfnamefont {A.}~\bibnamefont {Trautman}},\ }\bibfield  {title} {\enquote {\bibinfo {title} {{THE GEOMETRY OF GAUGE FIELDS. (TALK)}},}\ }\href {\doibase 10.1007/BF01603811} {\bibfield  {journal} {\bibinfo  {journal} {Czech. J. Phys. B}\ }\textbf {\bibinfo {volume} {29}},\ \bibinfo {pages} {107--116} (\bibinfo {year} {1979})}\BibitemShut {NoStop}%
\bibitem [{\citenamefont {Aad}\ \emph {et~al.}(2022{\natexlab{a}})\citenamefont {Aad} \emph {et~al.}}]{ATLAS:2022vkf}%
  \BibitemOpen
  \bibfield  {author} {\bibinfo {author} {\bibfnamefont {Georges}\ \bibnamefont {Aad}} \emph {et~al.} (\bibinfo {collaboration} {ATLAS}),\ }\bibfield  {title} {\enquote {\bibinfo {title} {{A detailed map of Higgs boson interactions by the ATLAS experiment ten years after the discovery}},}\ }\href {\doibase 10.1038/s41586-022-04893-w} {\bibfield  {journal} {\bibinfo  {journal} {Nature}\ }\textbf {\bibinfo {volume} {607}},\ \bibinfo {pages} {52--59} (\bibinfo {year} {2022}{\natexlab{a}})},\ \bibinfo {note} {[Erratum: Nature 612, E24 (2022)]},\ \Eprint {http://arxiv.org/abs/2207.00092} {arXiv:2207.00092 [hep-ex]} \BibitemShut {NoStop}%
\bibitem [{\citenamefont {Degrassi}\ \emph {et~al.}(2016)\citenamefont {Degrassi}, \citenamefont {Giardino}, \citenamefont {Maltoni},\ and\ \citenamefont {Pagani}}]{Degrassi:2016wml}%
  \BibitemOpen
  \bibfield  {author} {\bibinfo {author} {\bibfnamefont {Giuseppe}\ \bibnamefont {Degrassi}}, \bibinfo {author} {\bibfnamefont {Pier~Paolo}\ \bibnamefont {Giardino}}, \bibinfo {author} {\bibfnamefont {Fabio}\ \bibnamefont {Maltoni}}, \ and\ \bibinfo {author} {\bibfnamefont {Davide}\ \bibnamefont {Pagani}},\ }\bibfield  {title} {\enquote {\bibinfo {title} {{Probing the Higgs self coupling via single Higgs production at the LHC}},}\ }\href {\doibase 10.1007/JHEP12(2016)080} {\bibfield  {journal} {\bibinfo  {journal} {JHEP}\ }\textbf {\bibinfo {volume} {12}},\ \bibinfo {pages} {080} (\bibinfo {year} {2016})},\ \Eprint {http://arxiv.org/abs/1607.04251} {arXiv:1607.04251 [hep-ph]} \BibitemShut {NoStop}%
\bibitem [{\citenamefont {Lane}(2022)}]{Lane:2022ybv}%
  \BibitemOpen
  \bibfield  {author} {\bibinfo {author} {\bibfnamefont {Kenneth}\ \bibnamefont {Lane}},\ }\bibfield  {title} {\enquote {\bibinfo {title} {{The Composite Higgs Signal at the Next Big Collider}},}\ }in\ \href@noop {} {\emph {\bibinfo {booktitle} {{Snowmass 2021}}}}\ (\bibinfo {year} {2022})\ \Eprint {http://arxiv.org/abs/2203.03710} {arXiv:2203.03710 [hep-ph]} \BibitemShut {NoStop}%
\bibitem [{\citenamefont {Steingasser}\ and\ \citenamefont {Kaiser}(2023)}]{Steingasser:2023ugv}%
  \BibitemOpen
  \bibfield  {author} {\bibinfo {author} {\bibfnamefont {Thomas}\ \bibnamefont {Steingasser}}\ and\ \bibinfo {author} {\bibfnamefont {David~I.}\ \bibnamefont {Kaiser}},\ }\bibfield  {title} {\enquote {\bibinfo {title} {{Higgs potential criticality beyond the Standard Model}},}\ }\href {\doibase 10.1103/PhysRevD.108.095035} {\bibfield  {journal} {\bibinfo  {journal} {Phys. Rev. D}\ }\textbf {\bibinfo {volume} {108}},\ \bibinfo {pages} {095035} (\bibinfo {year} {2023})},\ \Eprint {http://arxiv.org/abs/2307.10361} {arXiv:2307.10361 [hep-ph]} \BibitemShut {NoStop}%
\bibitem [{\citenamefont {Aad}\ \emph {et~al.}(2024{\natexlab{a}})\citenamefont {Aad} \emph {et~al.}}]{ATLAS:2024ish}%
  \BibitemOpen
  \bibfield  {author} {\bibinfo {author} {\bibfnamefont {Georges}\ \bibnamefont {Aad}} \emph {et~al.} (\bibinfo {collaboration} {ATLAS}),\ }\bibfield  {title} {\enquote {\bibinfo {title} {{Combination of Searches for Higgs Boson Pair Production in pp Collisions at s=13\,\,TeV with the ATLAS Detector}},}\ }\href {\doibase 10.1103/PhysRevLett.133.101801} {\bibfield  {journal} {\bibinfo  {journal} {Phys. Rev. Lett.}\ }\textbf {\bibinfo {volume} {133}},\ \bibinfo {pages} {101801} (\bibinfo {year} {2024}{\natexlab{a}})},\ \Eprint {http://arxiv.org/abs/2406.09971} {arXiv:2406.09971 [hep-ex]} \BibitemShut {NoStop}%
\bibitem [{Dai(2019)}]{Dainese:2019rgk}%
  \BibitemOpen
  \href {\doibase /10.23731/CYRM-2019-007} {\enquote {\bibinfo {title} {{Report on the Physics at the HL-LHC and Perspectives for the HE-LHC}},}\ } (\bibinfo {year} {2019})\BibitemShut {NoStop}%
\bibitem [{\citenamefont {Chern}(1944)}]{Chern1944}%
  \BibitemOpen
  \bibfield  {author} {\bibinfo {author} {\bibfnamefont {Shiing-Shen}\ \bibnamefont {Chern}},\ }\bibfield  {title} {\enquote {\bibinfo {title} {A simple intrinsic proof of the gauss-bonnet formula for closed riemannian manifolds},}\ }\href {http://www.jstor.org/stable/1969302} {\bibfield  {journal} {\bibinfo  {journal} {Annals of Mathematics}\ }\textbf {\bibinfo {volume} {45}},\ \bibinfo {pages} {747--752} (\bibinfo {year} {1944})}\BibitemShut {NoStop}%
\bibitem [{\citenamefont {shen Chern}(1945)}]{Chern1945}%
  \BibitemOpen
  \bibfield  {author} {\bibinfo {author} {\bibfnamefont {Shiing}\ \bibnamefont {shen Chern}},\ }\bibfield  {title} {\enquote {\bibinfo {title} {On the curvatura integra in a riemannian manifold},}\ }\href {http://www.jstor.org/stable/1969203} {\bibfield  {journal} {\bibinfo  {journal} {Annals of Mathematics}\ }\textbf {\bibinfo {volume} {46}},\ \bibinfo {pages} {674--684} (\bibinfo {year} {1945})}\BibitemShut {NoStop}%
\bibitem [{\citenamefont {Coleman}\ and\ \citenamefont {Weinberg}(1973)}]{Coleman:1973jx}%
  \BibitemOpen
  \bibfield  {author} {\bibinfo {author} {\bibfnamefont {Sidney~R.}\ \bibnamefont {Coleman}}\ and\ \bibinfo {author} {\bibfnamefont {Erick~J.}\ \bibnamefont {Weinberg}},\ }\bibfield  {title} {\enquote {\bibinfo {title} {{Radiative Corrections as the Origin of Spontaneous Symmetry Breaking}},}\ }\href {\doibase 10.1103/PhysRevD.7.1888} {\bibfield  {journal} {\bibinfo  {journal} {Phys. Rev. D}\ }\textbf {\bibinfo {volume} {7}},\ \bibinfo {pages} {1888--1910} (\bibinfo {year} {1973})}\BibitemShut {NoStop}%
\bibitem [{\citenamefont {Rattazzi}\ and\ \citenamefont {Zaffaroni}(2001)}]{Rattazzi:2000hs}%
  \BibitemOpen
  \bibfield  {author} {\bibinfo {author} {\bibfnamefont {R.}~\bibnamefont {Rattazzi}}\ and\ \bibinfo {author} {\bibfnamefont {A.}~\bibnamefont {Zaffaroni}},\ }\bibfield  {title} {\enquote {\bibinfo {title} {{Comments on the holographic picture of the Randall-Sundrum model}},}\ }\href {\doibase 10.1088/1126-6708/2001/04/021} {\bibfield  {journal} {\bibinfo  {journal} {JHEP}\ }\textbf {\bibinfo {volume} {04}},\ \bibinfo {pages} {021} (\bibinfo {year} {2001})},\ \Eprint {http://arxiv.org/abs/hep-th/0012248} {arXiv:hep-th/0012248} \BibitemShut {NoStop}%
\bibitem [{\citenamefont {Ruegg}\ and\ \citenamefont {Ruiz-Altaba}(2004)}]{Ruegg:2003ps}%
  \BibitemOpen
  \bibfield  {author} {\bibinfo {author} {\bibfnamefont {Henri}\ \bibnamefont {Ruegg}}\ and\ \bibinfo {author} {\bibfnamefont {Marti}\ \bibnamefont {Ruiz-Altaba}},\ }\bibfield  {title} {\enquote {\bibinfo {title} {{The Stueckelberg field}},}\ }\href {\doibase 10.1142/S0217751X04019755} {\bibfield  {journal} {\bibinfo  {journal} {Int. J. Mod. Phys. A}\ }\textbf {\bibinfo {volume} {19}},\ \bibinfo {pages} {3265--3348} (\bibinfo {year} {2004})},\ \Eprint {http://arxiv.org/abs/hep-th/0304245} {arXiv:hep-th/0304245} \BibitemShut {NoStop}%
\bibitem [{\citenamefont {Capozziello}\ and\ \citenamefont {De~Laurentis}(2011)}]{Capozziello:2011et}%
  \BibitemOpen
  \bibfield  {author} {\bibinfo {author} {\bibfnamefont {Salvatore}\ \bibnamefont {Capozziello}}\ and\ \bibinfo {author} {\bibfnamefont {Mariafelicia}\ \bibnamefont {De~Laurentis}},\ }\bibfield  {title} {\enquote {\bibinfo {title} {{Extended Theories of Gravity}},}\ }\href {\doibase 10.1016/j.physrep.2011.09.003} {\bibfield  {journal} {\bibinfo  {journal} {Phys. Rept.}\ }\textbf {\bibinfo {volume} {509}},\ \bibinfo {pages} {167--321} (\bibinfo {year} {2011})},\ \Eprint {http://arxiv.org/abs/1108.6266} {arXiv:1108.6266 [gr-qc]} \BibitemShut {NoStop}%
\bibitem [{\citenamefont {Callan}\ \emph {et~al.}(1970)\citenamefont {Callan}, \citenamefont {Coleman},\ and\ \citenamefont {Jackiw}}]{Callan:1970ze}%
  \BibitemOpen
  \bibfield  {author} {\bibinfo {author} {\bibfnamefont {Curtis~G.}\ \bibnamefont {Callan}, \bibfnamefont {Jr.}}, \bibinfo {author} {\bibfnamefont {Sidney~R.}\ \bibnamefont {Coleman}}, \ and\ \bibinfo {author} {\bibfnamefont {Roman}\ \bibnamefont {Jackiw}},\ }\bibfield  {title} {\enquote {\bibinfo {title} {{A New improved energy - momentum tensor}},}\ }\href {\doibase 10.1016/0003-4916(70)90394-5} {\bibfield  {journal} {\bibinfo  {journal} {Annals Phys.}\ }\textbf {\bibinfo {volume} {59}},\ \bibinfo {pages} {42--73} (\bibinfo {year} {1970})}\BibitemShut {NoStop}%
\bibitem [{\citenamefont {Faraoni}(2004)}]{faraoniCosmologyScalartensorGravity2004}%
  \BibitemOpen
  \bibfield  {author} {\bibinfo {author} {\bibfnamefont {Valerio}\ \bibnamefont {Faraoni}},\ }\href@noop {} {\emph {\bibinfo {title} {Cosmology in Scalar-Tensor Gravity}}},\ \bibinfo {series} {Fundamental Theories of Physics}\ No.\ \bibinfo {number} {v. 139}\ (\bibinfo  {publisher} {Kluwer Academic Publishers},\ \bibinfo {address} {Dordrecht ; Boston},\ \bibinfo {year} {2004})\BibitemShut {NoStop}%
\bibitem [{\citenamefont {Aad}\ \emph {et~al.}(2024{\natexlab{b}})\citenamefont {Aad} \emph {et~al.}}]{ATLAS:2024erm}%
  \BibitemOpen
  \bibfield  {author} {\bibinfo {author} {\bibfnamefont {Georges}\ \bibnamefont {Aad}} \emph {et~al.} (\bibinfo {collaboration} {ATLAS}),\ }\bibfield  {title} {\enquote {\bibinfo {title} {{Measurement of the W-boson mass and width with the ATLAS detector using proton{\textendash}proton collisions at $\sqrt{s}=7$ TeV}},}\ }\href {\doibase 10.1140/epjc/s10052-024-13190-x} {\bibfield  {journal} {\bibinfo  {journal} {Eur. Phys. J. C}\ }\textbf {\bibinfo {volume} {84}},\ \bibinfo {pages} {1309} (\bibinfo {year} {2024}{\natexlab{b}})},\ \Eprint {http://arxiv.org/abs/2403.15085} {arXiv:2403.15085 [hep-ex]} \BibitemShut {NoStop}%
\bibitem [{\citenamefont {Aad}\ \emph {et~al.}(2023)\citenamefont {Aad} \emph {et~al.}}]{ATLAS:2023owm}%
  \BibitemOpen
  \bibfield  {author} {\bibinfo {author} {\bibfnamefont {Georges}\ \bibnamefont {Aad}} \emph {et~al.} (\bibinfo {collaboration} {ATLAS}),\ }\bibfield  {title} {\enquote {\bibinfo {title} {{Measurement of the Higgs boson mass with $H \to \gamma\gamma$ decays in 140 fb$^{-1}$ of $\sqrt{s}=13$ TeV $pp$ collisions with the ATLAS detector}},}\ }\href {\doibase 10.1016/j.physletb.2023.138315} {\bibfield  {journal} {\bibinfo  {journal} {Phys. Lett. B}\ }\textbf {\bibinfo {volume} {847}},\ \bibinfo {pages} {138315} (\bibinfo {year} {2023})},\ \Eprint {http://arxiv.org/abs/2308.07216} {arXiv:2308.07216 [hep-ex]} \BibitemShut {NoStop}%
\bibitem [{\citenamefont {Aad}\ \emph {et~al.}(2022{\natexlab{b}})\citenamefont {Aad} \emph {et~al.}}]{aad_detailed_2022}%
  \BibitemOpen
  \bibfield  {author} {\bibinfo {author} {\bibfnamefont {Georges}\ \bibnamefont {Aad}} \emph {et~al.} (\bibinfo {collaboration} {ATLAS}),\ }\bibfield  {title} {\enquote {\bibinfo {title} {{A detailed map of Higgs boson interactions by the ATLAS experiment ten years after the discovery}},}\ }\href {\doibase 10.1038/s41586-022-04893-w} {\bibfield  {journal} {\bibinfo  {journal} {Nature}\ }\textbf {\bibinfo {volume} {607}},\ \bibinfo {pages} {52--59} (\bibinfo {year} {2022}{\natexlab{b}})},\ \bibinfo {note} {[Erratum: Nature 612, E24 (2022)]},\ \Eprint {http://arxiv.org/abs/2207.00092} {arXiv:2207.00092 [hep-ex]} \BibitemShut {NoStop}%
\bibitem [{ATL(2025)}]{ATLAS-CONF-2025-006}%
  \BibitemOpen
  \href {https://cds.cern.ch/record/2937634} {\enquote {\bibinfo {title} {{Combined measurements of Higgs boson production and decay at $\sqrt{s} =$ 13 TeV using up to 140 fb$^{-1}$ of data collected by the ATLAS Experiment}},}\ } (\bibinfo {year} {2025})\BibitemShut {NoStop}%
\bibitem [{\citenamefont {Aad}\ \emph {et~al.}(2024{\natexlab{c}})\citenamefont {Aad} \emph {et~al.}}]{atlascollaboration2024measurementwbosonmasswidth}%
  \BibitemOpen
  \bibfield  {author} {\bibinfo {author} {\bibfnamefont {Georges}\ \bibnamefont {Aad}} \emph {et~al.} (\bibinfo {collaboration} {ATLAS}),\ }\bibfield  {title} {\enquote {\bibinfo {title} {{Measurement of the W-boson mass and width with the ATLAS detector using proton-proton collisions at $\sqrt{s}$ = 7 TeV}},}\ }\href@noop {} {\  (\bibinfo {year} {2024}{\natexlab{c}})},\ \Eprint {http://arxiv.org/abs/2403.15085} {arXiv:2403.15085 [hep-ex]} \BibitemShut {NoStop}%
\bibitem [{\citenamefont {Aad}\ \emph {et~al.}(2024{\natexlab{d}})\citenamefont {Aad} \emph {et~al.}}]{atlascollaboration2024characterisingHiggsbosonatlas}%
  \BibitemOpen
  \bibfield  {author} {\bibinfo {author} {\bibfnamefont {Georges}\ \bibnamefont {Aad}} \emph {et~al.} (\bibinfo {collaboration} {ATLAS}),\ }\bibfield  {title} {\enquote {\bibinfo {title} {{Characterising the Higgs boson with ATLAS data from Run 2 of the LHC}},}\ }\href {\doibase 10.1016/j.physrep.2024.11.001} {\bibfield  {journal} {\bibinfo  {journal} {Phys. Rept.}\ }\textbf {\bibinfo {volume} {11}},\ \bibinfo {pages} {001} (\bibinfo {year} {2024}{\natexlab{d}})},\ \Eprint {http://arxiv.org/abs/2404.05498} {arXiv:2404.05498 [hep-ex]} \BibitemShut {NoStop}%
\bibitem [{\citenamefont {Djouadi}(2008)}]{Djouadi:2005gi}%
  \BibitemOpen
  \bibfield  {author} {\bibinfo {author} {\bibfnamefont {Abdelhak}\ \bibnamefont {Djouadi}},\ }\bibfield  {title} {\enquote {\bibinfo {title} {{The Anatomy of electro-weak symmetry breaking. I: The Higgs boson in the standard model}},}\ }\href {\doibase 10.1016/j.physrep.2007.10.004} {\bibfield  {journal} {\bibinfo  {journal} {Phys. Rept.}\ }\textbf {\bibinfo {volume} {457}},\ \bibinfo {pages} {1--216} (\bibinfo {year} {2008})},\ \Eprint {http://arxiv.org/abs/hep-ph/0503172} {arXiv:hep-ph/0503172} \BibitemShut {NoStop}%
\bibitem [{\citenamefont {Chen}\ \emph {et~al.}(2016)\citenamefont {Chen}, \citenamefont {Yan}, \citenamefont {Zhao}, \citenamefont {Zhong},\ and\ \citenamefont {Zhao}}]{Chen:2015gva}%
  \BibitemOpen
  \bibfield  {author} {\bibinfo {author} {\bibfnamefont {Chien-Yi}\ \bibnamefont {Chen}}, \bibinfo {author} {\bibfnamefont {Qi-Shu}\ \bibnamefont {Yan}}, \bibinfo {author} {\bibfnamefont {Xiaoran}\ \bibnamefont {Zhao}}, \bibinfo {author} {\bibfnamefont {Yi-Ming}\ \bibnamefont {Zhong}}, \ and\ \bibinfo {author} {\bibfnamefont {Zhijie}\ \bibnamefont {Zhao}},\ }\bibfield  {title} {\enquote {\bibinfo {title} {{Probing triple-Higgs productions via 4b2\ensuremath{\gamma} decay channel at a 100 TeV hadron collider}},}\ }\href {\doibase 10.1103/PhysRevD.93.013007} {\bibfield  {journal} {\bibinfo  {journal} {Phys. Rev. D}\ }\textbf {\bibinfo {volume} {93}},\ \bibinfo {pages} {013007} (\bibinfo {year} {2016})},\ \Eprint {http://arxiv.org/abs/1510.04013} {arXiv:1510.04013 [hep-ph]} \BibitemShut {NoStop}%
\bibitem [{\citenamefont {Kilian}\ \emph {et~al.}(2017)\citenamefont {Kilian}, \citenamefont {Sun}, \citenamefont {Yan}, \citenamefont {Zhao},\ and\ \citenamefont {Zhao}}]{Kilian:2017nio}%
  \BibitemOpen
  \bibfield  {author} {\bibinfo {author} {\bibfnamefont {Wolfgang}\ \bibnamefont {Kilian}}, \bibinfo {author} {\bibfnamefont {Sichun}\ \bibnamefont {Sun}}, \bibinfo {author} {\bibfnamefont {Qi-Shu}\ \bibnamefont {Yan}}, \bibinfo {author} {\bibfnamefont {Xiaoran}\ \bibnamefont {Zhao}}, \ and\ \bibinfo {author} {\bibfnamefont {Zhijie}\ \bibnamefont {Zhao}},\ }\bibfield  {title} {\enquote {\bibinfo {title} {{New Physics in multi-Higgs boson final states}},}\ }\href {\doibase 10.1007/JHEP06(2017)145} {\bibfield  {journal} {\bibinfo  {journal} {JHEP}\ }\textbf {\bibinfo {volume} {06}},\ \bibinfo {pages} {145} (\bibinfo {year} {2017})},\ \Eprint {http://arxiv.org/abs/1702.03554} {arXiv:1702.03554 [hep-ph]} \BibitemShut {NoStop}%
\bibitem [{\citenamefont {Kilian}\ \emph {et~al.}(2020)\citenamefont {Kilian}, \citenamefont {Sun}, \citenamefont {Yan}, \citenamefont {Zhao},\ and\ \citenamefont {Zhao}}]{Kilian:2018bhs}%
  \BibitemOpen
  \bibfield  {author} {\bibinfo {author} {\bibfnamefont {Wolfgang}\ \bibnamefont {Kilian}}, \bibinfo {author} {\bibfnamefont {Sichun}\ \bibnamefont {Sun}}, \bibinfo {author} {\bibfnamefont {Qi-Shu}\ \bibnamefont {Yan}}, \bibinfo {author} {\bibfnamefont {Xiaoran}\ \bibnamefont {Zhao}}, \ and\ \bibinfo {author} {\bibfnamefont {Zhijie}\ \bibnamefont {Zhao}},\ }\bibfield  {title} {\enquote {\bibinfo {title} {{Multi-Higgs boson production and unitarity in vector-boson fusion at future hadron colliders}},}\ }\href {\doibase 10.1103/PhysRevD.101.076012} {\bibfield  {journal} {\bibinfo  {journal} {Phys. Rev. D}\ }\textbf {\bibinfo {volume} {101}},\ \bibinfo {pages} {076012} (\bibinfo {year} {2020})},\ \Eprint {http://arxiv.org/abs/1808.05534} {arXiv:1808.05534 [hep-ph]} \BibitemShut {NoStop}%
\bibitem [{\citenamefont {Goldberger}\ \emph {et~al.}(2008)\citenamefont {Goldberger}, \citenamefont {Grinstein},\ and\ \citenamefont {Skiba}}]{Goldberger:2007zk}%
  \BibitemOpen
  \bibfield  {author} {\bibinfo {author} {\bibfnamefont {Walter~D.}\ \bibnamefont {Goldberger}}, \bibinfo {author} {\bibfnamefont {Benjamin}\ \bibnamefont {Grinstein}}, \ and\ \bibinfo {author} {\bibfnamefont {Witold}\ \bibnamefont {Skiba}},\ }\bibfield  {title} {\enquote {\bibinfo {title} {{Distinguishing the Higgs boson from the dilaton at the Large Hadron Collider}},}\ }\href {\doibase 10.1103/PhysRevLett.100.111802} {\bibfield  {journal} {\bibinfo  {journal} {Phys. Rev. Lett.}\ }\textbf {\bibinfo {volume} {100}},\ \bibinfo {pages} {111802} (\bibinfo {year} {2008})},\ \Eprint {http://arxiv.org/abs/0708.1463} {arXiv:0708.1463 [hep-ph]} \BibitemShut {NoStop}%
\bibitem [{\citenamefont {Appelquist}\ \emph {et~al.}(2020)\citenamefont {Appelquist}, \citenamefont {Ingoldby},\ and\ \citenamefont {Piai}}]{appelquist_dilaton_2020}%
  \BibitemOpen
  \bibfield  {author} {\bibinfo {author} {\bibfnamefont {Thomas}\ \bibnamefont {Appelquist}}, \bibinfo {author} {\bibfnamefont {James}\ \bibnamefont {Ingoldby}}, \ and\ \bibinfo {author} {\bibfnamefont {Maurizio}\ \bibnamefont {Piai}},\ }\bibfield  {title} {\enquote {\bibinfo {title} {Dilaton potential and lattice data},}\ }\href {\doibase 10.1103/physrevd.101.075025} {\bibfield  {journal} {\bibinfo  {journal} {Physical Review D}\ }\textbf {\bibinfo {volume} {101}} (\bibinfo {year} {2020}),\ 10.1103/physrevd.101.075025}\BibitemShut {NoStop}%
\bibitem [{\citenamefont {Appelquist}\ \emph {et~al.}(2022)\citenamefont {Appelquist}, \citenamefont {Ingoldby},\ and\ \citenamefont {Piai}}]{appelquist_dilaton_2022}%
  \BibitemOpen
  \bibfield  {author} {\bibinfo {author} {\bibfnamefont {Thomas}\ \bibnamefont {Appelquist}}, \bibinfo {author} {\bibfnamefont {James}\ \bibnamefont {Ingoldby}}, \ and\ \bibinfo {author} {\bibfnamefont {Maurizio}\ \bibnamefont {Piai}},\ }\href {https://arxiv.org/abs/2209.14867} {\enquote {\bibinfo {title} {Dilaton effective field theory},}\ } (\bibinfo {year} {2022}),\ \Eprint {http://arxiv.org/abs/2209.14867} {arXiv:2209.14867 [hep-ph]} \BibitemShut {NoStop}%
\bibitem [{\citenamefont {Zwicky}(2024)}]{zwicky_qcd_2023}%
  \BibitemOpen
  \bibfield  {author} {\bibinfo {author} {\bibfnamefont {Roman}\ \bibnamefont {Zwicky}},\ }\href {https://arxiv.org/abs/2312.13761} {\enquote {\bibinfo {title} {Qcd with an infrared fixed point and a dilaton},}\ } (\bibinfo {year} {2024}),\ \Eprint {http://arxiv.org/abs/2312.13761} {arXiv:2312.13761 [hep-ph]} \BibitemShut {NoStop}%
\bibitem [{\citenamefont {Csaki}\ \emph {et~al.}(2007)\citenamefont {Csaki}, \citenamefont {Hubisz},\ and\ \citenamefont {Lee}}]{Csaki:2007ns}%
  \BibitemOpen
  \bibfield  {author} {\bibinfo {author} {\bibfnamefont {Csaba}\ \bibnamefont {Csaki}}, \bibinfo {author} {\bibfnamefont {Jay}\ \bibnamefont {Hubisz}}, \ and\ \bibinfo {author} {\bibfnamefont {Seung~J.}\ \bibnamefont {Lee}},\ }\bibfield  {title} {\enquote {\bibinfo {title} {{Radion phenomenology in realistic warped space models}},}\ }\href {\doibase 10.1103/PhysRevD.76.125015} {\bibfield  {journal} {\bibinfo  {journal} {Phys. Rev. D}\ }\textbf {\bibinfo {volume} {76}},\ \bibinfo {pages} {125015} (\bibinfo {year} {2007})},\ \Eprint {http://arxiv.org/abs/0705.3844} {arXiv:0705.3844 [hep-ph]} \BibitemShut {NoStop}%
\bibitem [{\citenamefont {He}\ \emph {et~al.}(2002)\citenamefont {He}, \citenamefont {Hill},\ and\ \citenamefont {Tait}}]{He_2002}%
  \BibitemOpen
  \bibfield  {author} {\bibinfo {author} {\bibfnamefont {Hong-Jian}\ \bibnamefont {He}}, \bibinfo {author} {\bibfnamefont {Christopher~T.}\ \bibnamefont {Hill}}, \ and\ \bibinfo {author} {\bibfnamefont {Tim M.~P.}\ \bibnamefont {Tait}},\ }\bibfield  {title} {\enquote {\bibinfo {title} {Top quark seesaw model, vacuum structure, and electroweak precision constraints},}\ }\href {\doibase 10.1103/physrevd.65.055006} {\bibfield  {journal} {\bibinfo  {journal} {Physical Review D}\ }\textbf {\bibinfo {volume} {65}} (\bibinfo {year} {2002}),\ 10.1103/physrevd.65.055006}\BibitemShut {NoStop}%
\bibitem [{\citenamefont {Huang}\ \emph {et~al.}(1990)\citenamefont {Huang}, \citenamefont {Wu},\ and\ \citenamefont {Zheng}}]{Huang:1989fj}%
  \BibitemOpen
  \bibfield  {author} {\bibinfo {author} {\bibfnamefont {Chao-guang}\ \bibnamefont {Huang}}, \bibinfo {author} {\bibfnamefont {Dan-di}\ \bibnamefont {Wu}}, \ and\ \bibinfo {author} {\bibfnamefont {Han-qing}\ \bibnamefont {Zheng}},\ }\bibfield  {title} {\enquote {\bibinfo {title} {{COSMOLOGICAL CONSTRAINTS TO WEYL'S VECTOR MESON}},}\ }\href@noop {} {\bibfield  {journal} {\bibinfo  {journal} {Commun. Theor. Phys.}\ }\textbf {\bibinfo {volume} {14}},\ \bibinfo {pages} {373--378} (\bibinfo {year} {1990})}\BibitemShut {NoStop}%
\bibitem [{\citenamefont {Tang}\ and\ \citenamefont {Wu}(2020)}]{Tang:2019uex}%
  \BibitemOpen
  \bibfield  {author} {\bibinfo {author} {\bibfnamefont {Yong}\ \bibnamefont {Tang}}\ and\ \bibinfo {author} {\bibfnamefont {Yue-Liang}\ \bibnamefont {Wu}},\ }\bibfield  {title} {\enquote {\bibinfo {title} {{Weyl Symmetry Inspired Inflation and Dark Matter}},}\ }\href {\doibase 10.1016/j.physletb.2020.135320} {\bibfield  {journal} {\bibinfo  {journal} {Phys. Lett. B}\ }\textbf {\bibinfo {volume} {803}},\ \bibinfo {pages} {135320} (\bibinfo {year} {2020})},\ \Eprint {http://arxiv.org/abs/1904.04493} {arXiv:1904.04493 [hep-ph]} \BibitemShut {NoStop}%
\bibitem [{\citenamefont {Ahriche}\ \emph {et~al.}(2024)\citenamefont {Ahriche}, \citenamefont {Kanemura},\ and\ \citenamefont {Tanaka}}]{ahriche2024gravitationalwavesphasetransitions}%
  \BibitemOpen
  \bibfield  {author} {\bibinfo {author} {\bibfnamefont {Amine}\ \bibnamefont {Ahriche}}, \bibinfo {author} {\bibfnamefont {Shinya}\ \bibnamefont {Kanemura}}, \ and\ \bibinfo {author} {\bibfnamefont {Masanori}\ \bibnamefont {Tanaka}},\ }\href {https://arxiv.org/abs/2308.12676} {\enquote {\bibinfo {title} {Gravitational waves from phase transitions in scale invariant models},}\ } (\bibinfo {year} {2024}),\ \Eprint {http://arxiv.org/abs/2308.12676} {arXiv:2308.12676 [hep-ph]} \BibitemShut {NoStop}%
\bibitem [{\citenamefont {Tang}\ and\ \citenamefont {Wu}(2018)}]{Tang_2018}%
  \BibitemOpen
  \bibfield  {author} {\bibinfo {author} {\bibfnamefont {Yong}\ \bibnamefont {Tang}}\ and\ \bibinfo {author} {\bibfnamefont {Yue-Liang}\ \bibnamefont {Wu}},\ }\bibfield  {title} {\enquote {\bibinfo {title} {Inflation in gauge theory of gravity with local scaling symmetry and quantum induced symmetry breaking},}\ }\href {\doibase 10.1016/j.physletb.2018.07.048} {\bibfield  {journal} {\bibinfo  {journal} {Physics Letters B}\ }\textbf {\bibinfo {volume} {784}},\ \bibinfo {pages} {163–168} (\bibinfo {year} {2018})}\BibitemShut {NoStop}%
\bibitem [{\citenamefont {Ghilencea}(2021)}]{Ghilencea_2021}%
  \BibitemOpen
  \bibfield  {author} {\bibinfo {author} {\bibfnamefont {D.~M.}\ \bibnamefont {Ghilencea}},\ }\bibfield  {title} {\enquote {\bibinfo {title} {Gauging scale symmetry and inflation: Weyl versus palatini gravity},}\ }\href {\doibase 10.1140/epjc/s10052-021-09226-1} {\bibfield  {journal} {\bibinfo  {journal} {The European Physical Journal C}\ }\textbf {\bibinfo {volume} {81}} (\bibinfo {year} {2021}),\ 10.1140/epjc/s10052-021-09226-1}\BibitemShut {NoStop}%
\bibitem [{\citenamefont {Donoghue}\ and\ \citenamefont {Menezes}(2018)}]{Donoghue:2017vvl}%
  \BibitemOpen
  \bibfield  {author} {\bibinfo {author} {\bibfnamefont {John~F.}\ \bibnamefont {Donoghue}}\ and\ \bibinfo {author} {\bibfnamefont {Gabriel}\ \bibnamefont {Menezes}},\ }\bibfield  {title} {\enquote {\bibinfo {title} {{Inducing the Einstein action in QCD-like theories}},}\ }\href {\doibase 10.1103/PhysRevD.97.056022} {\bibfield  {journal} {\bibinfo  {journal} {Phys. Rev. D}\ }\textbf {\bibinfo {volume} {97}},\ \bibinfo {pages} {056022} (\bibinfo {year} {2018})},\ \Eprint {http://arxiv.org/abs/1712.04468} {arXiv:1712.04468 [hep-ph]} \BibitemShut {NoStop}%
\bibitem [{\citenamefont {Migdal}\ and\ \citenamefont {Shifman}(1982)}]{migdal_dilaton_1982}%
  \BibitemOpen
  \bibfield  {author} {\bibinfo {author} {\bibfnamefont {A.A.}\ \bibnamefont {Migdal}}\ and\ \bibinfo {author} {\bibfnamefont {M.A.}\ \bibnamefont {Shifman}},\ }\bibfield  {title} {\enquote {\bibinfo {title} {Dilaton effective lagrangian in gluodynamics},}\ }\href {\doibase https://doi.org/10.1016/0370-2693(82)90089-2} {\bibfield  {journal} {\bibinfo  {journal} {Physics Letters B}\ }\textbf {\bibinfo {volume} {114}},\ \bibinfo {pages} {445--449} (\bibinfo {year} {1982})}\BibitemShut {NoStop}%
\bibitem [{\citenamefont {Weinberg}(1976)}]{Weinberg:1975gm}%
  \BibitemOpen
  \bibfield  {author} {\bibinfo {author} {\bibfnamefont {Steven}\ \bibnamefont {Weinberg}},\ }\bibfield  {title} {\enquote {\bibinfo {title} {{Implications of Dynamical Symmetry Breaking}},}\ }\href {\doibase 10.1103/PhysRevD.19.1277} {\bibfield  {journal} {\bibinfo  {journal} {Phys. Rev. D}\ }\textbf {\bibinfo {volume} {13}},\ \bibinfo {pages} {974--996} (\bibinfo {year} {1976})},\ \bibinfo {note} {[Addendum: Phys.Rev.D 19, 1277--1280 (1979)]}\BibitemShut {NoStop}%
\bibitem [{\citenamefont {Randall}\ and\ \citenamefont {Sundrum}(1999)}]{Randall:1999ee}%
  \BibitemOpen
  \bibfield  {author} {\bibinfo {author} {\bibfnamefont {Lisa}\ \bibnamefont {Randall}}\ and\ \bibinfo {author} {\bibfnamefont {Raman}\ \bibnamefont {Sundrum}},\ }\bibfield  {title} {\enquote {\bibinfo {title} {{A Large mass hierarchy from a small extra dimension}},}\ }\href {\doibase 10.1103/PhysRevLett.83.3370} {\bibfield  {journal} {\bibinfo  {journal} {Phys. Rev. Lett.}\ }\textbf {\bibinfo {volume} {83}},\ \bibinfo {pages} {3370--3373} (\bibinfo {year} {1999})},\ \Eprint {http://arxiv.org/abs/hep-ph/9905221} {arXiv:hep-ph/9905221} \BibitemShut {NoStop}%
\bibitem [{\citenamefont {Goldberger}\ and\ \citenamefont {Wise}(1999)}]{Goldberger:1999uk}%
  \BibitemOpen
  \bibfield  {author} {\bibinfo {author} {\bibfnamefont {Walter~D.}\ \bibnamefont {Goldberger}}\ and\ \bibinfo {author} {\bibfnamefont {Mark~B.}\ \bibnamefont {Wise}},\ }\bibfield  {title} {\enquote {\bibinfo {title} {{Modulus stabilization with bulk fields}},}\ }\href {\doibase 10.1103/PhysRevLett.83.4922} {\bibfield  {journal} {\bibinfo  {journal} {Phys. Rev. Lett.}\ }\textbf {\bibinfo {volume} {83}},\ \bibinfo {pages} {4922--4925} (\bibinfo {year} {1999})},\ \Eprint {http://arxiv.org/abs/hep-ph/9907447} {arXiv:hep-ph/9907447} \BibitemShut {NoStop}%
\bibitem [{\citenamefont {Frampton}\ and\ \citenamefont {Vafa}(1999)}]{Frampton:1999yb}%
  \BibitemOpen
  \bibfield  {author} {\bibinfo {author} {\bibfnamefont {Paul~H.}\ \bibnamefont {Frampton}}\ and\ \bibinfo {author} {\bibfnamefont {Cumrun}\ \bibnamefont {Vafa}},\ }\bibfield  {title} {\enquote {\bibinfo {title} {{Conformal approach to particle phenomenology}},}\ }\href@noop {} {\  (\bibinfo {year} {1999})},\ \Eprint {http://arxiv.org/abs/hep-th/9903226} {arXiv:hep-th/9903226} \BibitemShut {NoStop}%
\bibitem [{\citenamefont {Csaki}\ \emph {et~al.}(2016)\citenamefont {Csaki}, \citenamefont {Grojean},\ and\ \citenamefont {Terning}}]{Csaki:2015hcd}%
  \BibitemOpen
  \bibfield  {author} {\bibinfo {author} {\bibfnamefont {Csaba}\ \bibnamefont {Csaki}}, \bibinfo {author} {\bibfnamefont {Christophe}\ \bibnamefont {Grojean}}, \ and\ \bibinfo {author} {\bibfnamefont {John}\ \bibnamefont {Terning}},\ }\bibfield  {title} {\enquote {\bibinfo {title} {{Alternatives to an Elementary Higgs}},}\ }\href {\doibase 10.1103/RevModPhys.88.045001} {\bibfield  {journal} {\bibinfo  {journal} {Rev. Mod. Phys.}\ }\textbf {\bibinfo {volume} {88}},\ \bibinfo {pages} {045001} (\bibinfo {year} {2016})},\ \Eprint {http://arxiv.org/abs/1512.00468} {arXiv:1512.00468 [hep-ph]} \BibitemShut {NoStop}%
\bibitem [{\citenamefont {Cs{\'a}ki}\ \emph {et~al.}(2018)\citenamefont {Cs{\'a}ki}, \citenamefont {Lombardo},\ and\ \citenamefont {Telem}}]{Csaki:2018muy}%
  \BibitemOpen
  \bibfield  {author} {\bibinfo {author} {\bibfnamefont {Csaba}\ \bibnamefont {Cs{\'a}ki}}, \bibinfo {author} {\bibfnamefont {Salvator}\ \bibnamefont {Lombardo}}, \ and\ \bibinfo {author} {\bibfnamefont {Ofri}\ \bibnamefont {Telem}},\ }\bibfield  {title} {\enquote {\bibinfo {title} {{TASI lectures on non-supersymmetric BSM models.}}}\ \ }(\bibinfo  {publisher} {WSP},\ \bibinfo {year} {2018})\ pp.\ \bibinfo {pages} {501--570},\ \Eprint {http://arxiv.org/abs/1811.04279} {arXiv:1811.04279 [hep-ph]} \BibitemShut {NoStop}%
\bibitem [{\citenamefont {Cs{\'a}ki}\ \emph {et~al.}(2022)\citenamefont {Cs{\'a}ki}, \citenamefont {Lombardo},\ and\ \citenamefont {Telem}}]{Csaki:2021abz}%
  \BibitemOpen
  \bibfield  {author} {\bibinfo {author} {\bibfnamefont {Csaba}\ \bibnamefont {Cs{\'a}ki}}, \bibinfo {author} {\bibfnamefont {Salvator}\ \bibnamefont {Lombardo}}, \ and\ \bibinfo {author} {\bibfnamefont {Ofri}\ \bibnamefont {Telem}},\ }\bibfield  {title} {\enquote {\bibinfo {title} {{Non-supersymmetric BSM models}},}\ }\href {\doibase 10.1140/epjs/s11734-021-00343-2} {\bibfield  {journal} {\bibinfo  {journal} {Eur. Phys. J. ST}\ }\textbf {\bibinfo {volume} {231}},\ \bibinfo {pages} {1229--1264} (\bibinfo {year} {2022})}\BibitemShut {NoStop}%
\bibitem [{\citenamefont {Graham}\ \emph {et~al.}(2015)\citenamefont {Graham}, \citenamefont {Kaplan},\ and\ \citenamefont {Rajendran}}]{Graham:2015cka}%
  \BibitemOpen
  \bibfield  {author} {\bibinfo {author} {\bibfnamefont {Peter~W.}\ \bibnamefont {Graham}}, \bibinfo {author} {\bibfnamefont {David~E.}\ \bibnamefont {Kaplan}}, \ and\ \bibinfo {author} {\bibfnamefont {Surjeet}\ \bibnamefont {Rajendran}},\ }\bibfield  {title} {\enquote {\bibinfo {title} {{Cosmological Relaxation of the Electroweak Scale}},}\ }\href {\doibase 10.1103/PhysRevLett.115.221801} {\bibfield  {journal} {\bibinfo  {journal} {Phys. Rev. Lett.}\ }\textbf {\bibinfo {volume} {115}},\ \bibinfo {pages} {221801} (\bibinfo {year} {2015})},\ \Eprint {http://arxiv.org/abs/1504.07551} {arXiv:1504.07551 [hep-ph]} \BibitemShut {NoStop}%
\bibitem [{\citenamefont {Ferretti}(2022)}]{Ferretti:2021jai}%
  \BibitemOpen
  \bibfield  {author} {\bibinfo {author} {\bibfnamefont {Gabriele}\ \bibnamefont {Ferretti}},\ }\bibfield  {title} {\enquote {\bibinfo {title} {{Underlying gauge-fermion models of compositeness}},}\ }\href {\doibase 10.1140/epjs/s11734-021-00217-7} {\bibfield  {journal} {\bibinfo  {journal} {Eur. Phys. J. ST}\ }\textbf {\bibinfo {volume} {231}},\ \bibinfo {pages} {1265--1272} (\bibinfo {year} {2022})}\BibitemShut {NoStop}%
\bibitem [{\citenamefont {Cat{\`a}}\ \emph {et~al.}(2019)\citenamefont {Cat{\`a}}, \citenamefont {Crewther},\ and\ \citenamefont {Tunstall}}]{Cata:2018wzl}%
  \BibitemOpen
  \bibfield  {author} {\bibinfo {author} {\bibfnamefont {O.}~\bibnamefont {Cat{\`a}}}, \bibinfo {author} {\bibfnamefont {R.~J.}\ \bibnamefont {Crewther}}, \ and\ \bibinfo {author} {\bibfnamefont {Lewis~C.}\ \bibnamefont {Tunstall}},\ }\bibfield  {title} {\enquote {\bibinfo {title} {{Crawling technicolor}},}\ }\href {\doibase 10.1103/PhysRevD.100.095007} {\bibfield  {journal} {\bibinfo  {journal} {Phys. Rev. D}\ }\textbf {\bibinfo {volume} {100}},\ \bibinfo {pages} {095007} (\bibinfo {year} {2019})},\ \Eprint {http://arxiv.org/abs/1803.08513} {arXiv:1803.08513 [hep-ph]} \BibitemShut {NoStop}%
\bibitem [{\citenamefont {Crewther}(2020)}]{Crewther:2020tgd}%
  \BibitemOpen
  \bibfield  {author} {\bibinfo {author} {\bibfnamefont {R.~J.}\ \bibnamefont {Crewther}},\ }\bibfield  {title} {\enquote {\bibinfo {title} {{Genuine Dilatons in Gauge Theories}},}\ }\href {\doibase 10.3390/universe6070096} {\bibfield  {journal} {\bibinfo  {journal} {Universe}\ }\textbf {\bibinfo {volume} {6}},\ \bibinfo {pages} {96} (\bibinfo {year} {2020})},\ \Eprint {http://arxiv.org/abs/2003.11259} {arXiv:2003.11259 [hep-ph]} \BibitemShut {NoStop}%
\end{thebibliography}%

\end{document}